\documentclass[a4paper,11pt]{article}
\pdfoutput=1
\usepackage{jcappub}
\usepackage[T1]{fontenc} 
\usepackage{newtxtext,newtxmath}

\title{A physically-motivated template set for high-$z$ galaxy SED fitting}

\author[a,1]{Judah Luberto,\note{Corresponding author.}}
\author[a]{Steven Furlanetto,}
\author[b, c]{Jordan Mirocha.}

\affiliation[a]{Department of Physics and Astronomy, University of California, Los Angeles, CA 90024, USA}
\affiliation[b]{Jet Propulsion Laboratory, California Institute of Technology, 4800 Oak Grove Drive, Pasadena, CA 91109, USA}
\affiliation[c]{California Institute of Technology, 1200 E. California Boulevard, Pasadena, CA 91125, USA}

\emailAdd{judah@astro.ucla.edu}

\abstract{We introduce a new physically-motivated spectral template set for fitting the spectral energy distributions (SEDs) of high-$z$ galaxies. We use the public galaxy formation code \textsc{ares} to generate star formation histories of thirteen representative galaxies with diverse masses and generate their predicted spectra across a set of redshifts at $z > 6$. The model parameters are calibrated to reproduce the properties of $z>6$ galaxies observed by HST. Motivated by the apparent importance of bursty star formation at high redshifts, we also include templates with recent starbursts. We use these templates with the SED-fitting code EAZY to analyze both an independent theoretical model and a public sample of JWST-observed galaxies from the JADES survey. The comparison with a semi-analytic model demonstrates that our fitting framework accurately measures the galaxy properties, even when the underlying assumptions of the model differ from ours. Our preliminary application to JWST data shows that galaxies at $z > 8$ are often bursty (especially at small galaxy masses), follow a star-forming main sequence similar to those at lower redshift (albeit with a higher normalization), and form stars earlier than expected in \textsc{ares}. Our SED-fitting framework is very fast (thanks to the efficiency of EAZY) but provides full inferred star formation histories for each source. Additionally, it enables a direct comparison to theoretical models and helps point toward improvements necessary in those models.}


\begin{document}
\maketitle
\flushbottom

\section{Introduction}

The current hubbub in the high-redshift ($z > 6$) galaxy-formation community is how the \textit{James Webb Space Telescope} (JWST) will improve our understanding of galaxy formation during the first billion years of the Universe's history. Due to its wavelength coverage and sensitivity, it will observe the highest-$z$ galaxies to date. For the first time, we hope to understand star- and galaxy-formation models during the Cosmic Dawn --- which is important because a number of unanswered questions (e.g., the timescale of the Epoch of Reionization, the population-scale growth of early galaxies) can be constrained with observations at the highest redshifts \citep{Bouwens2015, Mason2015}.

Galaxy-formation models have so far explained properties of $z \lesssim 6$ galaxy populations (such as the luminosity and stellar mass functions) by way of simple physical principles like star-formation regulation from stellar feedback (e.g., \cite{Mason2015, Gentry2017, Hopkins2018a, Kawakatu2020}). These principles should extend to higher redshifts (although doubt has recently been cast on this assumption \cite{Greene2023, Labbe2023b, Matthee2023}), but with some important differences: at high-$z$, galaxies live in smaller potential wells that grow more rapidly, for example. Models following these principles have produced reasonable fits to HST data \citep{Somerville2015, Furlanetto2017}, and we can now compare them to the new JWST results.

As the first JWST observations of high-redshift galaxies rolled in, large discrepancies were found between observations and the models \citep{Robertson2023, Labbe2023a, BoylanKolchin2023, ArrabalHaro2023, Lovell2023}, particularly with an overabundance of bright galaxies at high redshifts. While spectroscopic followup of these candidate galaxies have shown some of them to be interlopers (due to the similar dropout appearance in blue filters between a high-$z$ Lyman-break galaxy and a low-$z$ dusty red galaxy; \cite{Desprez2023, Fujimoto2023, Kocevski2023, Rodighiero2023, Zavala2023, Glazebrook2023}), many others have been confirmed \citep{Curtis-Lake2023, Haro2023, Wang2023, Looser2023}. 

A suite of papers followed that worked to remedy this disagreement between models and observations. Some found that the bright galaxies can be explained with a higher star formation efficiency and/or more scatter in the star formation rate at a fixed mass (``burstiness'') \citep{Dekel2023, Mason2023, Mirocha&Furlanetto2023, Munoz2023, Sun2023b, Yung2024}. Others explored dust evolution \citep{Ferrera2023, Ferrara2024}, AGN contributions \citep{Hegde&Wyatt2024}, and even alternative cosmologies (e.g., \citep{liu_accelerating_2022, gupta_jwst_2023, steinhardt_highest_2023, adil_dark_2023, padmanabhan_alleviating_2023, menci_negative_2024}).  

Determining which of these solutions are physically relevant requires a detailed understanding of these sources, which means probing their physical properties. The primary tool to do this is spectral energy distribution (SED) fitting. This process takes photometry for each source and fits a stellar population to it, from which properties like the redshift, star formation rate, stellar mass, metallicity, dust extinction, etc. can be measured. Of course, even more detailed information can be extracted from spectroscopic measurements, but such observations are much more expensive than photometric surveys so are limited to a smaller number of sources.

A number of individual codes perform SED fitting --- each creating the ``best-fit'' spectra in their own way. Many of these codes (e.g., \textsc{beagle} \cite{Chevallard2016}, \textsc{bagpipes} \cite{Carnall2018}, \textsc{prospector} \cite{Johnson2021}, \textsc{ProSpect} \cite{Robotham2020}) alter a star formation history (SFH) to match features in the observed spectra. Some of these impose a SFH functional form to do the tuning, while others take a non-parameterized approach, where the user sets bins in time during which the star formation rate (SFR) is constant, and the code modulates the amplitude of the SFR within these bins. 

However, SED fitting contains a number of subtle challenges. A crucial concern is that the parameterized SFHs must assume a sensible functional form to get a reasonable fit while the non-paramaterized SFHs must use physically-appropriate bins of constant SFR \citep{Lower2020, Jain2024}. The quality of these fits can also suffer from recent bursts in star formation \citep{Haskell2023, Narayanan2024}, the assumed IMF shape \citep{Wang2024}, or incorrect dust properties \citep{Zavala2023}; there too may be a bias between photometrically-fitted results and spectroscopic results \citep{Duan2023, Wang2024}.

Unfortunately, these problems are particularly acute for another SED-fiting code, Easy and Accurate Z$_{\mathrm{phot}}$ from Yale (EAZY) \citep{Brammer2008}, which otherwise has tremendous advantages --- especially in its speed and ease of use. EAZY  uses a small set of rest spectra of galaxies called a ``template set'' that are carefully chosen to ``span'' the galaxy population of interest. The code determines the best-fitting linear combination of templates, from which the galaxy properties (e.g., redshift, SFR, stellar mass, dust content) can be estimated as well. For surveys of the early Universe, EAZY is particularly popular for estimating source redshifts (e.g., \cite{Conselice2024, Finkelstein2024, Arribas2024, Weibel2024}).

Because many JWST-observed galaxies are at higher redshifts than any seen before, there is an understandable worry about applying the standard EAZY templates to high-$z$ JWST galaxies. This set does remarkably well in fitting low-redshift galaxies, but it was not designed to describe their high-$z$ counterparts. This has led some to ``fill in'' EAZY with more templates appropriate to (known) high-$z$ galaxies \citep{Hainline2023, Larson2023}. But these phenomenological approaches lack physical information about the templates so they cannot be used to measure physical properties beyond the galaxy redshift. These issues have led many to use EAZY only for redshift fitting and not for estimating the properties of the galaxies.

In this paper, we take a different approach by using the galaxy-formation code \textsc{ares} to create a physically-motivated template set specific to high-$z$ galaxies. With this template set, we know by construction the full properties of each galaxy (e.g., SFH, stellar mass, etc.), and thus can infer the properties of the observed sources (at least in the context of this theoretical model), and not just the redshift. Because the template set is made of high-$z$ model galaxies, the results are both more credible and  open an avenue for direct comparisons of observations and models. Enabling these kinds of inferences inside EAZY's lightning-fast analysis framework will accelerate the discovery process of JWST surveys and ensure that as much information as possible can be squeezed from the data (which still rely on photometric measurements for the vast majority of sources). We will show that EAZY  is a good tool for this endeavor: its efficient template sets make pulling physical information easy.

The paper is structured as follows. In Section~\ref{sec:EAZY}, we introduce the EAZY framework for creating a template set, and in Section~\ref{sec:ares} we create the template set itself using the \textsc{ares} code. Section~\ref{sec:fittingscsam} validates our templates by comparing them to a completely independent theoretical model. Finally, in Section~\ref{sec:fitjwstdata} we compare to data from the recent JADES survey. We conclude and suggest some next steps in Section~\ref{sec:concl}.

We adopt adopt cosmological parameters very similar to the recent \cite{Planck2020} constraints: $\Omega_{m} = 0.3156$, $\Omega_{b} = 0.0491$, $h = 0.6726$, and $\sigma_{8} = 0.8159$.

\section{SED-Fitting with EAZY} \label{sec:EAZY}

Here we provide a brief overview of the structure of EAZY; we refer the interested reader to \cite{Brammer2008} for more details. EAZY's strength is in its small template set and simple structure, which allows for very rapid fitting and parameter inference. With high-$z$ JWST surveys, it has been used primarily to estimate redshifts for sources, but it can also estimate the source properties. Given a template set, EAZY executes the following steps to fit a galaxy SED:

\begin{enumerate}
    \item For each template, produce photometric points matching the filters used in the observations by integrating the template spectrum over the transmission curves of the filters. This results in a set of photometric points for each template spectrum: an $(N, M)$ array with $N$ templates and $M$ filters. 
    
    \item Create a ``set'' of photometric points by redshifting each template's photometry to each $z$ over a user-specified range (for instance, vary $z$ across the range [0, 20] with $\Delta z = 0.01$). The array is now of shape $(N, M, K)$ with $K$ being the length of the redshift array. As a default, emission blueward of Ly$\alpha$ is assumed to be completely absorbed by the IGM at the redshifts relevant to our study. 

    While typical users of EAZY populate this series by redshifting a single template, EAZY includes functionality to create separate \textit{instances} of the rest spectra at each redshift (essentially changing the underlying template as a function of redshift). Thus when creating the $(N, M, K)$ array, the rest spectrum instance closest to each redshift is used. Our approach uses this functionality, as we will explain later in Section~\ref{sec:zdep}.

    \item For each redshift in the range, find the linear combination of template spectra which best fits the galaxy's photometry. Assign this best fit a $\chi^{2}$ value --- a goodness of fit. The maximum likelihood redshift corresponds to the fit with the minimum $\chi^2$, and its distribution can be used to construct the posterior of the source redshift, $P(z)$.
    
    \item If the templates have known properties (e.g., stellar mass, star formation rate, luminosities, dust content), use the linear coefficients associated with each template at the maximum-likelihood redshift to infer the galaxy's properties.
\end{enumerate}
 EAZY's strength in speed is also its weakness. It puts its full faith in the accuracy of the user-chosen template sets. If the templates are incomplete (or even incorrect), the fit will be biased. This is a prominent problem in high redshift space where galaxy formation is not well constrained. 

In the low-redshift range, where galaxy formation is better understood, EAZY has done quite well. Its original default template set was carefully chosen to be both compact and to span a broad range of galaxies, including passively evolving and star-forming types, and it has been ``tweaked'' over the years to match new observations better. However, the utility of this set for high redshifts is questionable. Theoretical models suggest that the vast majority of $z > 6$ galaxies will be forming stars rapidly, with very young, blue stellar populations. EAZY's default template set has three spectra with discernable Lyman-limit breaks. Only one of those templates has a negative F200W - F277W color (is blue) \citep{Larson2023}. While EAZY does remove all flux blueward of the Ly$\alpha$ break for high-$z$ sources, the lack of templates suited for high-$z$ fits can give poor fits. We will show some specific examples of these difficulties in later sections.

This worry of ill-fitting has led many to only use EAZY for redshift fitting (which relies mostly on the Lyman break so is fairly robust to the details of the templates), ignoring the parameter outputs. Previous attempts at creating high redshift template sets for EAZY have focused only on redshift fitting, either trying to fix the color incompleteness in the EAZY default templates \citep{Larson2023} or supplementing the default templates with rest spectra that fit galaxies from high-$z$ simulations well \citep{Hainline2023, Steinhardt2023}.

Another worry with EAZY's template set (and SED-fitting in general) also affects the redshift inference: if the template set is incomplete, it will necessarily miss redshift solutions from the unknown population. For high-$z$ galaxies, this is a particular concern for old, dusty, and/or line-emitting galaxies from lower redshifts \citep{Furlanetto&Mirocha2023, Desprez2023, Fujimoto2023, Kocevski2023, Rodighiero2023, Zavala2023, Glazebrook2023}, where the Lyman break can be confused with other spectral breaks. 

This paper's goal is to tend to the physical inferences problem by introducing a template set tuned to high-$z$ galaxies. We do not attempt to model interlopers from lower redshifts. 

\section{Using \textsc{ares} To Construct a Template Set} \label{sec:ares}

In this section, we will construct a template set for EAZY using a flexible model of high-redshift galaxies. Such a template set will both better estimate the parameters of high-redshift galaxies (as the precise details of the model-generated templates are known) and test the efficacy of the model itself.

At the end of this process, we will have a template set consisting of 13 galaxies with a variety of physically-motivated star formation histories. Each template is ``redshift dependent'' (Section \ref{sec:zdep}) and has the possibility of a burst component containing up to $33 \%$ of its stellar mass at that redshift (Section \ref{sec:burst}).

The final templates are available at \url{https://github.com/JudahRockLuberto/high-z_template_set}.

\subsection{Star Formation History Modeling with \textsc{ares}}

We begin by creating a set of galaxies using the Accelerated Reionization Era Simulations (\textsc{ares}) code\footnote{In particular, we use the ``v1'' branch at \url{https://github.com/mirochaj/ares}.}. \textsc{ares} is a public code which takes dark matter halo growth histories and pairs them with a phenomenological model for star formation. The model assumes pristine gas from the intergalactic medium accretes onto the halo and forms stars. The halo growth histories are generated assuming halos grow at constant number density \citep{Furlanetto2017}, using the  \cite{Tinker2010} halo mass function\footnote{\textsc{ares} uses \texttt{HMFcalc} \citep{Murray2013} to calculate the halo mass function.}. The star formation is modulated by an efficiency, $f_{*}$, that depends on the halo mass $M_h$ and (optionally) the redshift:
\begin{equation}
    \dot{M}_{*} (M_{h}, z) = f_{*} (M_{h}, z) \dot{M}_{b} (M_{h}, z)
\end{equation}
where $\dot{M}_{*}$ is the SFR, $f_{*}$ is the star formation efficiency (SFE), and $\dot{M}_{b}$ is the baryonic mass accretion rate. The star formation efficiency is a double power-law in $M_{h}$ with the form
\begin{equation}
    f_{*} (M_{h}) = \frac{f_{*, 10} C_{10}}{\left( M_{h}/M_{p} \right)^{- \alpha_{*, lo}} + \left( M_{h}/M_{p} \right)^{- \alpha_{*, hi}}}
\end{equation}
where $M_{p}$ is the location of the peak, $\alpha_{*, hi}$ and $\alpha_{*, lo}$ are the power-law slopes on either side of the peak, and $f_{*, 10}$ and $C_{10}$ are normalizing constants, referring to the SFE at $10^{10} M_{\odot}$ and $C_{10} \equiv ( 10^{10} / M_{p} )^{- \alpha_{*, lo}} + ( 10^{10} / M_{p} )^{- \alpha_{*, hi}}$, respectively. 

Once the star formation history for a halo is known, \textsc{ares} generates the intrinsic spectrum by summing the stellar contributions from each generation of stars using the BPASSv1 code \citep{Eldridge2009}. The spectrum is supplemented by \cite{Ferland1980} for the nebular continuum, while the BPASSv1 code has nebular line emissions built in. The metallicity is fixed across stellar mass and redshift at $Z = 0.004$.

Dust is included by assuming the galaxy sits in the center of a uniform, spherically-symmetric dust cloud. The optical depth of such a cloud is
\begin{equation}
    \tau_{\lambda} = \kappa_{\lambda} \frac{3 M_{d}}{4 \pi R_{d}^{2}}
\end{equation}
where $\kappa_{\lambda}$ is the absorption coefficient, $M_{d}$ is the dust mass, and $R_{d}$ is a characteristic dust scale length. The absorption coefficient is a power-law consistent with the dust power-law of the SMC \citep{Weingartner2001}. The dust scale length is a double power-law with the mass of the halo.

One of the strengths of \textsc{ares} is its flexibility. For example, the star formation efficiency can be prescribed using physical models of stellar feedback (e.g. energy- and/or momentum-driven supernova \citep{Dekel2014, Furlanetto2017}, or empirically calibrated to galaxy luminosity functions and colors \citep{Sun2016, Tacchella2018, Behroozi2019, Mirocha2020}). \textsc{ares} can also shift its default prescription for dust in the interstellar medium (e.g. the rate at which metals get lodged in the ISM). It also includes a prescription for imposing stochastic variations in the star formation rate by drawing from a log-normal distribution about the mean, although we do not use that here.

For this work, we ran an \textsc{ares} instance using the model from \cite{Mirocha2020}, which has been shown to reproduce the rest-ultraviolet (UV) luminosity function and UV colors up to $z \sim 10$ using HST data. The assumed parameter values are in Table 1 of \cite{Mirocha2020}. While this parameter set does not reproduce higher-redshift UVLFs measured by JWST \citep{Mirocha&Furlanetto2023}, we will find that it produces a sufficiently flexible template set to provide good fits to high-$z$ galaxies. In future work, we will explore how \textsc{ares} galaxy formation parameters can be inferred from observations by optimizing the template set.

\subsection{Creating a Set of \textsc{ares} galaxies}

\begin{figure}
    \centering
    \includegraphics[width=\textwidth]{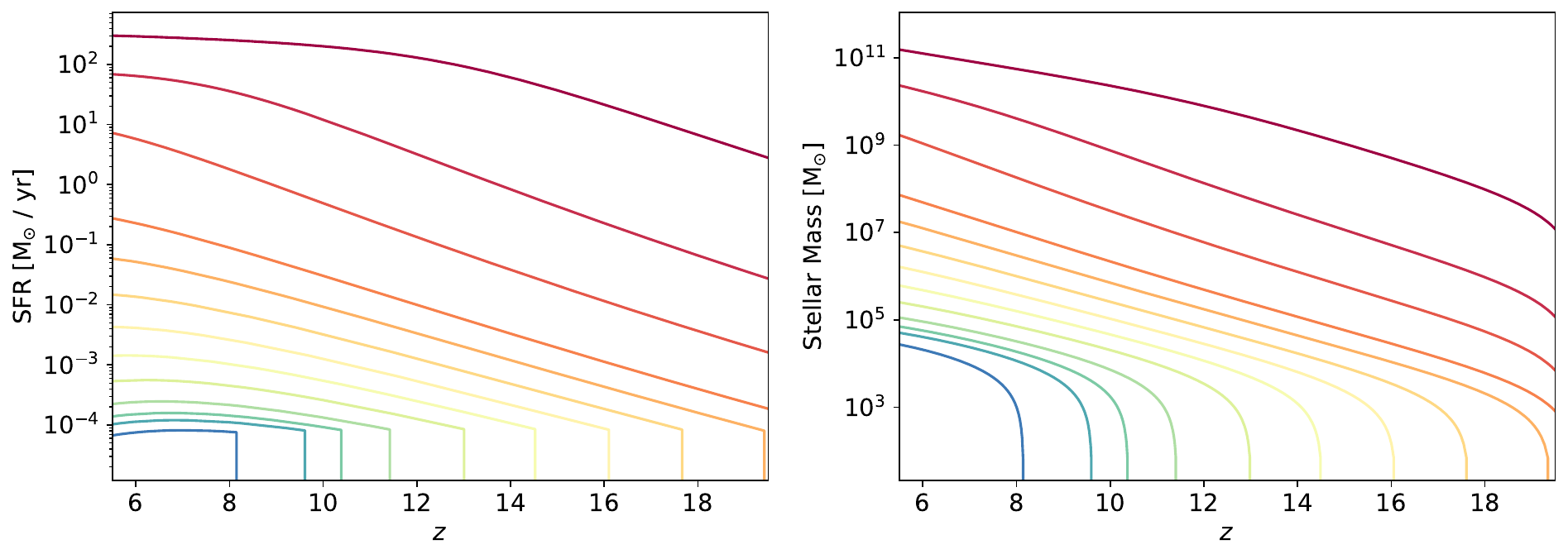}
    \caption{The star formation histories (left) and stellar mass growth histories (right) of our template galaxies as a function of redshift. The colors are the same in both panels for each galaxy; at $z=6$, the total halo masses range from $\sim10^{8}$--$10^{12} \; \mathrm{M_{\odot}}$. Creating a complete template set with a variety of star formation histories is necessary to obtain high quality fits, considering that in \textsc{ares} low-mass galaxies grow differently than high-mass galaxies, as do high and low-redshift galaxies. The templates span $z = 6$--$19$.}
    \label{fig:sfh-mass}
\end{figure}

Using the parameters of \cite{Mirocha2020}, we can create a small template set of a diverse collection of galaxies. One way to do this is to choose galaxies with a wide range of star formation histories. Motivated by this rule, we selected 13 galaxies which began star formation at varying times: six of these began star formation with a spacing of $\Delta z \sim 1.5$ from $z \approx 10 - 20$, four began star formation earlier than $z = 20$, and three were added which began star formation at $z \approx 8, 9.5, 10.5$ based on the general prior that more galaxies formed at lower redshift than at higher redshift. Because our halos grow monotonically (and smoothly), these initial times correspond to specific halo masses at any given redshift: at $z=6$, they range from $\sim10^{8}$--$10^{12} \; \mathrm{M_{\odot}}$.

Figure \ref{fig:sfh-mass} shows their star formation histories in the left panel and the corresponding stellar mass histories in the right panel. 
Note that because some of these galaxies do not form stars until $z < 10$, we are effectively fitting with fewer templates at higher redshifts. This follows the trend of higher redshifts having less variety in galaxy formation. It also marginally decreases the fitting time at higher redshifts.

Figure~\ref{fig:sfh-mass} shows that galaxies are expected to grow very rapidly through the Cosmic Dawn in \textsc{ares}. This is typical of most models of high-$z$ galaxy formation (e.g., \cite{Dayal2013, Mason2015, Furlanetto2017, Tacchella2018}), although of course we expect real galaxies to have scatter around these simple models due to mergers and other complications. The SFRs of these galaxies result from the combination of their accretion rates (which increase monotonically with halo mass) and the star formation efficiency (which is fit to observations, peaking at moderate halo masses before declining). We note that the full set of templates provides a wide range in overall ages and star formation rates.

\begin{figure}
    \centering
    \includegraphics[width=\textwidth]{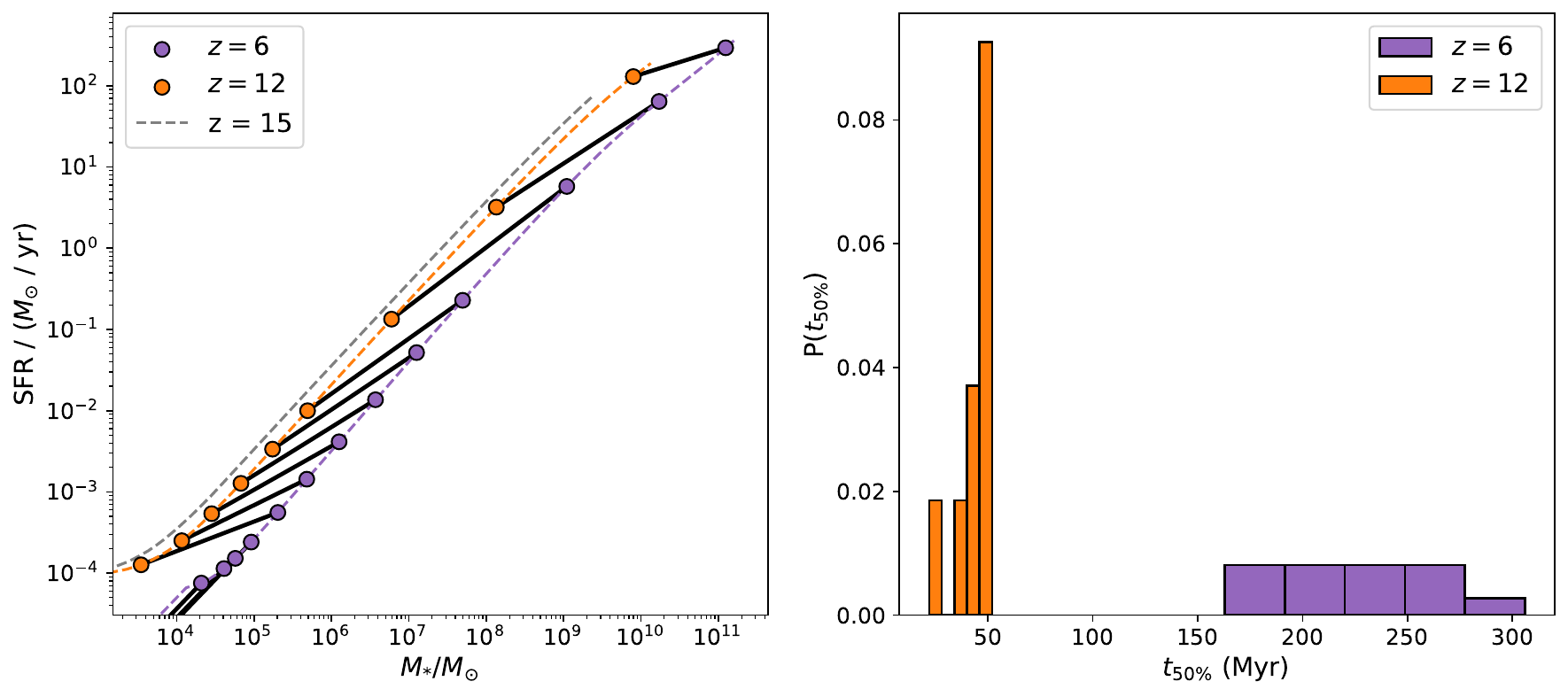}
    \caption{\textit{Left:} 
    The star-forming galaxy main sequence (i.e., star formation rate vs. stellar mass) for the \textsc{ares} galaxies at $z = 6$, $z = 12$, and $z = 15$, in purple, orange, and gray dashed lines, respectively. The dots show the location of each template galaxy at $z=6$ and 12 (the bottommost purple dots correspond to galaxies that have not formed stars before $z = 12$), and the solid lines between the dots indicate the paths the galaxies evolve. Because galaxies at higher redshifts have less time to reach a fixed stellar mass than their lower-$z$ counterparts, they must form stars more quickly to reach that stellar mass, and thus have higher SFRs. Therefore, the star-forming main sequence evolves across redshifts, suggesting that fitting galaxies at one redshift from templates from another will result in errors. \textit{Right:} Distribution of mean stellar ages, defined as the half-mass time, or how long ago $50\%$ of stars were formed in each galaxy, for the $z = 6$ and $z = 12$ template galaxies.}
    \label{fig:template_properties}
\end{figure}

Figure~\ref{fig:template_properties} further explores the properties of these model galaxies. In the left panel, we show the range of stellar masses and star formation rates at $z = 6$ (purple) and $z = 12$ (orange). The loci containing these points define the star-forming galaxy main sequence at these redshifts (shown by the dashed curves); we also show this for $z=15$ for reference. We note that this main sequence evolves rapidly over this redshift range, which implies that the specific star formation rate evolves significantly in the model galaxies. This is qualitatively consistent with the evolving main sequence at lower redshifts \citep{Schreiber2015, Pearson2018, Popesso2023} and with early measurements from JWST at $z > 6$ \citep{Clarke2024, Roberts-Borsani2024}. This has direct implications for SED fitting templates: qualitatively, we expect more UV light compared to the optical for early galaxies. 

We also note that our template set includes very small systems --- some with stellar masses $\lesssim~10^4~\ M_\odot$. Such systems would be far too faint to be visible in conventional galaxy surveys, but we emphasize that the fitting procedure allows the overall stellar mass to vary. These systems have very recent star formation, so their templates will contribute when a (larger) observed galaxy is dominated by recent star formation (though also see section~\ref{sec:burst} below). 

The distribution of mean stellar ages at $z = 6$ and $z = 12$ are also shown in the right panel of figure \ref{fig:template_properties}. Here we show the lookback time $t_{50\%}$ (measured from the ``observed'' redshift) during which 50\% of the galaxy's stars formed. This is simply a reflection of the growth timescale of halos in a $\Lambda$CDM cosmology, which (roughly) scales with the Hubble time. At a fixed stellar mass, galaxies at $z=12$ have much less time to build up than galaxies at $z=6$, so they must have smaller ages. This has important implications for SED-fitting with templates: the template set must include a wide range of formation timescales in order to span the rapidly-changing galaxies during the Cosmic Dawn.

\begin{figure}
    \centering
    \includegraphics[width=\textwidth]{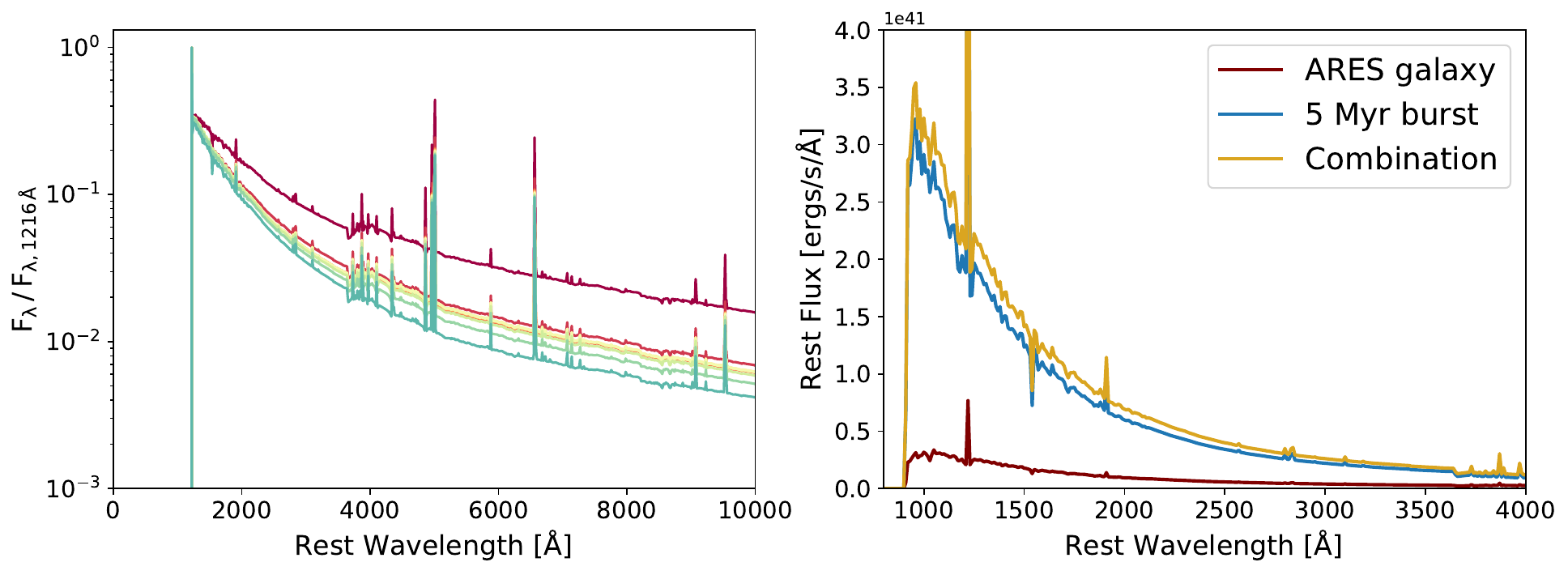}
    \caption{\textit{Left:} The rest spectra of each template at $z = 9$, normalized to the Lyman-$\alpha$ break at $\lambda = 1216 \ \text{\AA}$. The colors match the galaxies in figure \ref{fig:sfh-mass}, except the most blue object in figure \ref{fig:sfh-mass} is not included here because it has not yet begun forming stars at $z = 9$. The varying spectral shape across halo masses point to the need for a variety of high-$z$ spectra in SED fitting. Note that we have assumed the IGM absorbs all light blueward of Ly$\alpha$ here (as EAZY assumes for high-$z$ galaxies).
    \textit{Right:} An example rest spectrum of a galaxy in our template set (red) with the inclusion of a SSP burst 5 Myr old (gold) which
    makes up $33\%$ of the mass of the template. The burst spectrum is isolated in blue. Note that these spectra have Ly$\alpha$ and the Lyman continuum included.}
    \label{fig:spectra-and-burst}
\end{figure}

The rest spectra at $z = 9$ for each of these galaxies is shown in (the left panel of) figure \ref{fig:spectra-and-burst}, normalized to the Lyman-$\alpha$ break at $\lambda = 1216 \ \text{\AA}$. Each spectrum is colored according to the halo mass, as in figure~\ref{fig:sfh-mass}, although the most blue SFH in figure \ref{fig:sfh-mass} has not started forming stars yet and is not included in the figure. The unique spectral shape for each galaxy supports the need for a template set that spans a range of star formation histories. It is hard to imagine a linear combination of low-mass spectra reproducing a high-mass spectra. 

\subsection{A Redshift-Dependent Template Set} \label{sec:zdep}

In the previous section, we found that \textsc{ares} galaxies vary across two important dimensions: \emph{(i)} at a fixed redshift, massive galaxies have redder spectra than low-mass galaxies (see figure~\ref{fig:spectra-and-burst}) and \emph{(ii)} at fixed stellar mass, stars form more rapidly at higher redshifts (as exemplified by the lack of overlap of the $z = 6$ vs. $z = 12$ distributions in figure~\ref{fig:template_properties}). This implies that --- even if the templates at each redshift are accurate --- we cannot simply choose a set of model spectra at a single redshift and expect them to fit across the entire era. For example, suppose a source ($z \sim 10$) is fit with templates from an incorrect redshift ($z \sim 6$). Because $z \sim 6$ galaxies have slightly more dust than $z \sim 10$ galaxies, the SFR we infer for the $z \sim 10$ galaxy based on the rest-UV emission will be biased high. Similarly, for the stellar mass, because the $z \sim 6$ galaxy will have had more time to form stars compared to the $z \sim 10$ galaxy, the stellar masses will be biased high. We show this explicitly in section \ref{sec:param_estimation}.

To remedy these difficulties, we rely on a largely unused ability in EAZY: template sets can be redshift dependent. As mentioned in section \ref{sec:EAZY}, a redshift-dependent template set fits an input galaxy at a redshift $z$ with the template spectrum instances that are nearest to that $z$. 

We thus choose to make our template set redshift dependent. We take the chosen galaxies from \textsc{ares}, calculate their rest spectra across the interval $z = [6,19]$, spaced every $\Delta z = 0.1$, so that each template has 131 spectra associated with it, although we delay their inclusion into the template set until 10 Myr \textit{after} the galaxies begin forming stars, because \textsc{ARES} does not attempt to model the beginning of star formation accurately. We note that, despite the increase in the number of stored spectra, the computation time increases only slightly, because no more than 13 spectra are used for any one fit. (Indeed, at high redshifts some of the template galaxies have not formed yet, so the fit is performed on \emph{fewer} templates.)

The redshift-dependent templates also offer an additional advantage: because they are drawn from a single \textsc{ares} parameter set, they can be used to test the redshift evolution assumed by that parameter combination. We do not take advantage of this constraining power in this work, but it is a useful tool for the future.

\subsection{A Template Set with Burstiness} \label{sec:burst}

There is now strong evidence that burstiness is a key component in low-mass, high-$z$ star formation, regulating the growth of the galaxies \citep{Muratov2015, Facher-Giguere2018, Furlanetto2022}. Observations too have found evidence of burstiness in high-$z$ galaxies, using photometry \citep{Ciesla2024, Dressler2023a, Dressler2023b, Endsley2023}
and spectroscopy \citep{Caputi2017, Faisst2019, Rinaldi2022, Looser2023}, and it may be important in explaining the excess of bright galaxies at $z > 10$ \citep{Mason2023, Mirocha&Furlanetto2023, Sun2023b, Dekel2023}. 

\textsc{ares} includes an optional stochastic component to mimic burstiness. But that is not ideal for creating a template set, as it would imprint particular stochastic fluctuations onto each template galaxy. We therefore take another strategy. 

Often nonparametric SED fitting codes include a bin for recent star formation that is only a few Myr long whose amplitude can be modulated to represent recent bursts. A simple analog of this would be to introduce a new template of the spectrum of a recent burst,  including it along with the \textsc{ares} models. However, EAZY does not have the native capability to include a prior on the contribution of any one template (and hence the size of the burst), so that it could be fit with $100 \%$ of the stellar mass, which should be extremely rare. 

We have implemented a workaround: in addition to our 13 smoothly-evolving \textsc{ares} galaxies, we have added another group of 13 templates that include contributions from the same galaxies and a dust-free star burst that begins 5~Myr before the observation, declining exponentially, with a 1~Myr exponential decay, much like the results of models and simulations \citep{Emami2019}. We chose 5~Myr because the UV emission is dominated by young, massive stars which live on the order of $\sim$$1 - 10$~Myr, so 5~Myr is a reasonable average age for the burst (also see section~\ref{sec:systematics}).

This approach effectively sets a prior on the maximum burst of star formation because each galaxy's burst and non-burst counterpart can be linearly combined to produce a burst ranging from $0$--$33\%$ of the total stellar mass (where in our case we have chosen $33\%$ to be the maximum, or, the burst mass can reach $50 \%$ of the underlying \textsc{ares} galaxy mass) at each redshift. Figure \ref{fig:spectra-and-burst} is a visualization of this process (right panel). We show the rest spectrum at $z=8$ of an example \textsc{ares} galaxy in red, as well as its maximum burst (comprising $33 \%$ of the template's stellar mass at $z=8$) in blue. The combination of the two (which is the corresponding burst template) is shown in gold. In total at each redshift, we have a set of 13 galaxies like the red spectrum and a set of 13 galaxies like the gold spectrum.

While this approach initially seems to double the number of galaxies in our template set, we are really only adding one degree of freedom to the fitting (a burst component). This particular implementation is necessary only to allow us to enforce a maximum prior on a single starburst. We also note that we do not attempt to add additional templates to model older bursts, because once the contribution from massive star fades they would be difficult to distinguish from other old stars produced by the normal templates. We will show below that many observed galaxies prefer these recent burst templates. 

\subsection{The Ly$\alpha$ Line}

Our spectral templates assume that ionizing photons from massive stars are absorbed within the galaxies and reprocessed into line radiation, including Ly$\alpha$. However, to reach the observer these photons must navigate not only the complex environment of the galaxy (including dust and winds) but also the IGM, which is expected to be extremely optically thick to Ly$\alpha$ photons through most of this era (see, e.g., \cite{Ouchi2020} and references therein). Ly$\alpha$ lines will only survive if their source galaxies are embedded in large ionized bubbles. All these make Ly$\alpha$'s absorption from high-redshift galaxies highly uncertain, dependent on the properties of the galaxy itself and the surrounding environment.

Handling Ly$\alpha$ is thus tricky when comparing data to our models. If our template set makes incorrect assumptions about Ly$\alpha$, the redshift fit and properties can wildly differ from the true values. 

To be as agnostic as possible, we therefore create two separate template sets: one that includes the intrinsic Ly$\alpha$ emission and one that assumes it is fully absorbed. The templates without Ly$\alpha$ are the exact same as the templates with it, except we assume Ly$\alpha$ is totally absorbed by the surrounding neutral gas and replace the signal with a continuum. In particular, we take the average flux (in the rest frame) of the two pixels adjacent to the Ly$\alpha$ signal, both of which are outside the edges of the line. 

We note that these two sets are left separate (unlike the additional burst templates). In this work, we will mostly use the the ``no Ly$\alpha$'' set, under the assumption that most of the Ly$\alpha$ lines will be absorbed during reionization. Our public release includes both versions, so that users can choose how to treat Ly$\alpha$. 

\subsection{Qualitative Comparison to Other SED-fitting Approaches} \label{sec:full-sfh-model}

We have already seen that EAZY is just one of many SED-fitting codes used to estimate the parameters of observed galaxies. These codes are powerful indeed: while the stellar mass, SFR, and redshift are obvious quantities, these only scratch the surface --- as embedded in the spectrum is information about the growth of the galaxy across cosmic time. The spectrum is ultimately built from the sum of individual star's contributions, where each star's SED evolves as a function of it's mass and age. Therefore, encoded in a spectrum is the entire star formation history of the galaxy. This information is crucial for ``cosmic archaeology'' --- using systems observed at one time to learn about star formation earlier in the Universe's history. For instance, early JWST results have found an over-abundance of luminous high-$z$ galaxies, and to explore this surprise, many have modeled the SFH of early JWST galaxies \citep{Dressler2023a, 
Qin2023, Whitler2023} to pin down when the galaxies formed these unexpectedly large stellar populations.

To obtain so much information, however, a reasonable guess at the shape of the galaxy's SFH is necessary. Usually, SED fitting codes  either use a user-defined parameterized SFH, adjusting the parameters to obtain the best fit (e.g., \textsc{beagle} \cite{Chevallard2016}, \textsc{Bagpipes} \cite{Carnall2018}), or scale user-defined time bins of constant star formation (e.g., \textsc{Prospector} \cite{Johnson2021}, \textsc{ProSpect} \cite{Robotham2020}) to find the best fit. 

EAZY does not natively have this capability because it finds the best fits by scaling and summing its template spectra, rather than by adjusting an assumed SFH shape. Therefore, one of EAZY's weaknesses in SED fitting is that it does not directly model a full SFH but instead reports the contribution of each template to the final spectrum. 

We overcome this challenge by leveraging the well-known SFHs in each of our \textsc{ares} model templates to approximate the SFH of the fits. To get the full SFH from EAZY, we use the linear coefficients associated with each fit. These coefficients can be directly mapped to a full star formation history: we take the dot product of the EAZY linear coefficients and the SFHs associated with our templates, and sum them up, for a complete estimated SFH.

In comparison to other SED-fitting codes, this approach has both advantages and disadvantages. The most obvious limitation is that the shape of the model SFHs is fixed for each template. This is qualitatively similar to SED-fitting codes with parameterized SFHs: they put a lot of faith is put in the functional form, while we put a lot of faith in our templates. Our approach results in less transparent input histories but relates more directly to the expectations of modeling. We also note that we are not allowing the model parameters to vary, so an effective prior is put on the range of allowed SFHs. However, we show later that the  shape is not as rigid as it may seem, and there is enough flexibility in the templates to match a range of galaxies. 

There are two obvious advantages to our approach. The first is speed, a hallmark of the EAZY code. The second is that our parameters have clear physical meanings. An important aspect of this is that cosmological evolution's role in building a stellar population is built into the model. Our template galaxies incorporate redshift evolution in order to connect observations across a range of redshifts with one underlying model. In the future, we can use JWST measurements to directly constrain these parameters. However, we do note that because we use only a single galaxy model in our template set, we cannot yet constrain variations in some parameters (like the metallicity). We will see that these do not have much effect on photometric surveys, but presumably such variations would be important in comparisons to spectroscopic surveys.

\section{Applying the Templates to an Independent Theory Model} \label{sec:fittingscsam}

The purpose of this section is to show that our templates provide a reasonable fit to a completely independent model. The Santa Cruz semi-analytic model (SAM) \citep{Somerville1999, Somerville2001, Somerville2015} is a well-established physically-motivated galaxy-formation model. Using stellar population synthesis (SPS) and full star formation histories, mock lightcones and catalogs were created, tuned to the view of five CANDELS surveys \citep{Yung2022}. We use this published simulated galaxy catalog as validation for our template set, and to harness the extended abilities of our template set to make general comparisons between the Santa Cruz SAM and the \textsc{ares} model. 

However, it is important to note some key differences between the Santa Cruz SAM and the \textsc{ares} model. Namely, the Santa Cruz SAM uses the \cite{BC03} model for stellar population synthesis (SPS), a Chabrier initial mass function (IMF) \citep{Chabrier2003}, and the Padova1994 stellar isochrones \citep{Bertelli1994}. 
While the \textsc{ares} model uses the Chabrier IMF, it uses \textsc{BPASS} version 1.0 \citep{Eldridge2009} for SPS. These differences can change the fitting results slightly, which should be considered when interpreting our fits.

To perform this comparison in a way that mimics observations, we imagine the Santa Cruz SAM model as a mock survey, using a number of wide and medium filters in the mid- and near-infrared. Similar to \cite{Larson2023}, we set the photometric errors to the $1 \sigma$ CEERS depth of $m=30.72$ \citep{Finklestein2017}, and we impose a $\mathrm{S/N > 3}$ restriction on this catalog, cutting any galaxies below this baseline. After this cut, there are $\sim$900,000 galaxies between $z = 0 - 10$ and $\sim$3000 above $z > 8$. We choose the $z > 8$ galaxies for the analysis, which is useful because we can ignore potential interlopers, knowing we are fitting high-$z$ model galaxies.

To create the most balanced analysis possible, we use our template set without Ly$\alpha$, because \cite{Yung2022} did not include Ly$\alpha$ emission.

\subsection{Redshift Estimation With Our Template Set}
 
We now compare our results with two recent template sets calibrated for high redshifts \cite{Hainline2023, Larson2023}, and the EAZY default template set (called "tweak\_fsps\_QSF\_12\_v3.param"). In particular, we use the ``tweak\_fsps\_QSF\_12\_v3\_newtemplates.param'' template from \cite{Larson2023}, which is for fitting galaxies in $z > 8$. We make this comparison because it is important to ensure our template set performs as well as others in a fundamental property like redshift.

\begin{figure}
    \centering
    \includegraphics[width=0.6\textwidth]{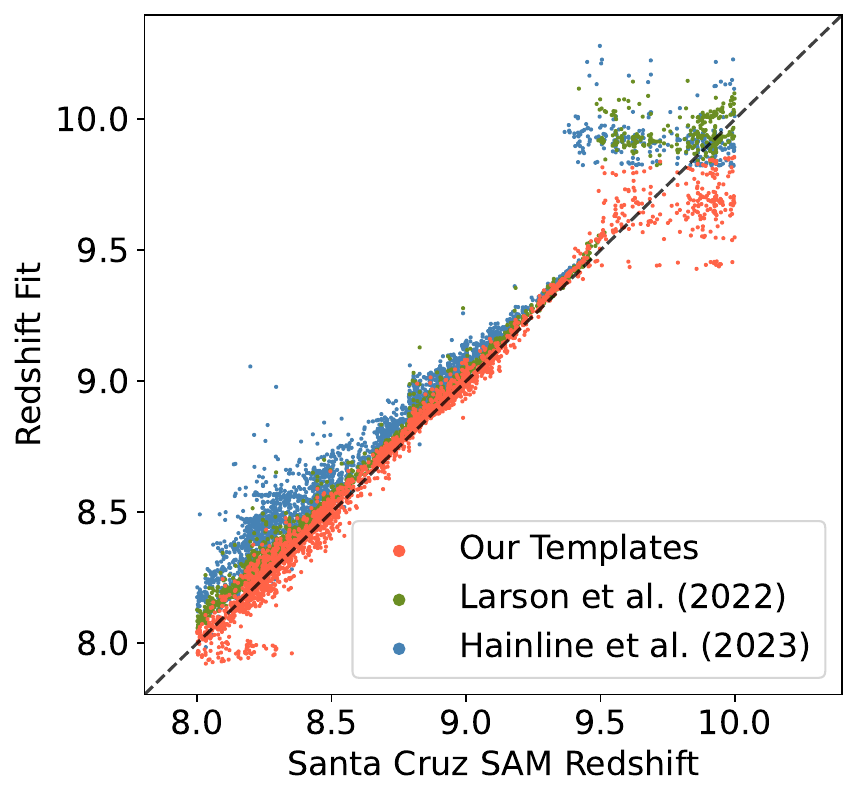}
    \caption{True redshifts of SC SAM z > 8 galaxies against the redshift fitting from our template set (orange), from \cite{Larson2023} (green), and from \cite{Hainline2023} (blue). The agreement between the three template sets is good; all accurately reproduce the input redshifts within $\Delta z \sim 0.1$, although our templates appear to be somewhat less biased. Note that the poor fits at $z \sim 10$ are due to a small gap in JWST filter coverage for the Lyman break at this redshift.}
    \label{fig:SC-SAM-z}
\end{figure}

We show the redshift-fit results in figure \ref{fig:SC-SAM-z}. The results are approximately equal between each template set made for high-redshift galaxies, although the \cite{Larson2023} and \cite{Hainline2023} templates are biased slightly high ($\Delta z \sim 0.1$). Note that we do not show the redshift fits from the EAZY default set because many of the inferred values were at low redshift ($z < 3$), preferring a fit with an old and red galaxy template instead of a Ly$\alpha$ break template. The comparable performance of the three sets of high-$z$ templates is reassuring; we do not read too much into the lower redshift bias in our framework because neither \textsc{ares} nor the Santa Cruz SAM matches high-$z$ galaxies perfectly.

With each template set, there is a strip of ill-fitting $z$ at the upper end of the redshift range too. This is likely due to the gap between the F115W and F150W NIRCam filters at $~ 1.2 \mathrm{\mu m}$, corresponding with the Ly$\alpha$ break at $z \approx 9.5$--$10$. This is a redshift range where it is difficult to pin down redshifts well.

\begin{figure}
    \centering
    \includegraphics[width=0.6\textwidth]{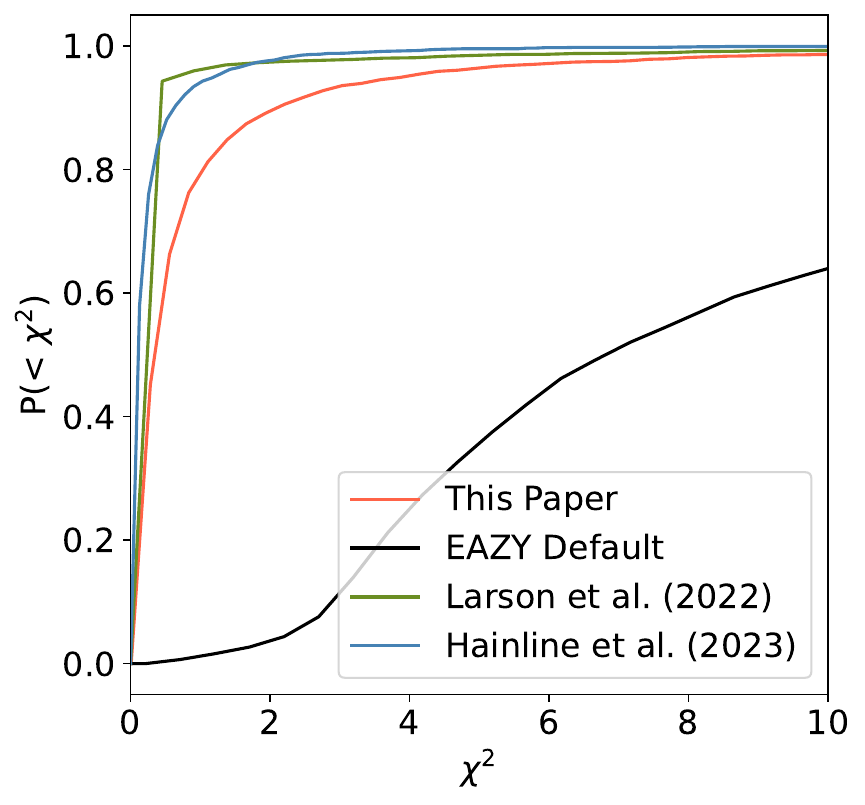}
    \caption{$\chi^{2}$ cumulative distributions from these fits for each template set, including EAZY's default template set ("tweak\_fsps\_QSF\_12\_v3.param") in black. The different template sets qualititatively have the same $\chi^{2}$ distribution (or quality of fits).}
    \label{fig:SC-SAM_chi2}
\end{figure}

It is prudent to check the $\chi^{2}$ distributions of the fits, which measures the quality of each fit. We plot the distributions of $\chi^2$ for each template set in figure \ref{fig:SC-SAM_chi2}. We make a $\chi^{2} < 10$ cut for viewability, although clearly the EAZY default set extends to much larger values (and hence poorer fits). 

Unfortunately, it is difficult to make firm conclusions beyond the fact that the other template sets appear to fit the photometry well. A rigorous comparison between the template sets is hard: in particular, the number of free parameters in each one is unknown. The EAZY default-template-set galaxies (which both \cite{Larson2023} and \cite{Hainline2023} are built on) are meant to be ``principal-component-like'' so that each template is a totally different kind of galaxy. Each of these may add a degree of freedom. However, the \textsc{ares} galaxy spectra are covariant as the SFHs follow a model with its own parameters. Also, the number of \textsc{ares} galaxies available for fitting decreases with increasing redshift. For a fit of a general galaxy, the ``more accurate'' template set of the three compared here is not obvious by looking at the $\chi^{2}$s. Thus we simply conclude that all the template sets do a reasonable job fitting the high-$z$ Santa Cruz SAM galaxies. The advantage of our template set, however, is the ability to pull physical information from the fits.

\subsection{Measuring Galaxy Properties With Our Template Set} \label{sec:param_estimation}

The quality of redshift fits is only one component of the result; the fitted physical parameters are also crucial to understanding galaxy evolution. One way to confirm that our framework is robust is to test it with galaxies of known parameters, which the Santa Cruz SAM has, and compare the results with both the EAZY default template set and more simplified versions of our template set. This will demonstrate that the physical processes incorporated into our full model are necessary and sufficient to match a broad swath of theoretical models.

\begin{figure}
    \centering
    \includegraphics[width=0.6\textwidth]{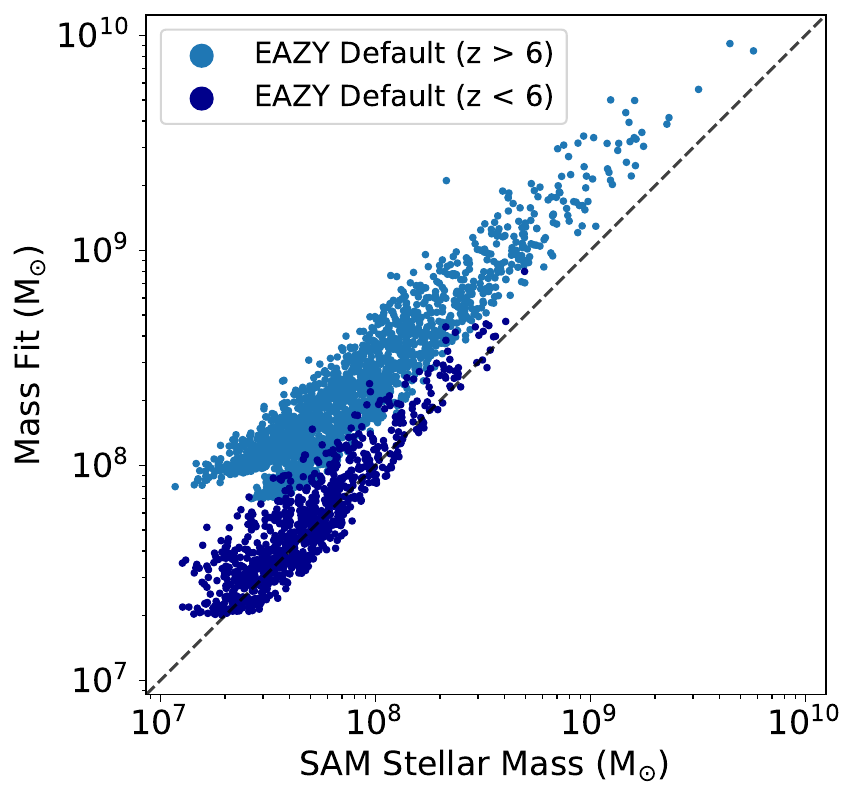}
    \caption{The inferred stellar masses of Santa Cruz SAM galaxies with the EAZY default template set. The fits are split into galaxies fit as $z < 6$ (dark blue) and $z > 6$ (light blue). The EAZY fits at $z > 6$, while within the correct range of redshifts, significantly overestimate the stellar masses. Oddly, fits that (incorrectly) place the galaxies at $z<6$ roughly match the model stellar masses, although this is not helpful given the code misidentifies the galaxy. These kinds of incorrect results are difficult to trust and support the need for an improved template set.}
    \label{fig:SC-SAM-EAZY-only}
\end{figure}

To motivate the need for a new template set, figure \ref{fig:SC-SAM-EAZY-only} compares the stellar mass fits from the EAZY-default template set to the known Santa Cruz SAM values. We divide EAZY's default template set values into fits dominated by their Ly$\alpha$ break galaxy (which result in relatively accurate redshifts at $z>6$; light blue) and by all others (which result in incorrect redshifts at $z<3$; dark blue). Notice how the stellar mass values for the $z > 6$ dominated fits are biased high by up to an order of magnitude. This is a well-known issue with EAZY fitting and one reason why EAZY has frequently been used for redshift fitting only \citep{Finkelstein2022, Naidu2022, Nelson2023, Robertson2023, Duan2024, Rinaldi2024}. The galaxies being fit as $z < 6$ sources have poor fits overall, although they do straddle the ``correct fit'' line.

\begin{figure}
    \centering
    \includegraphics[width=\textwidth]{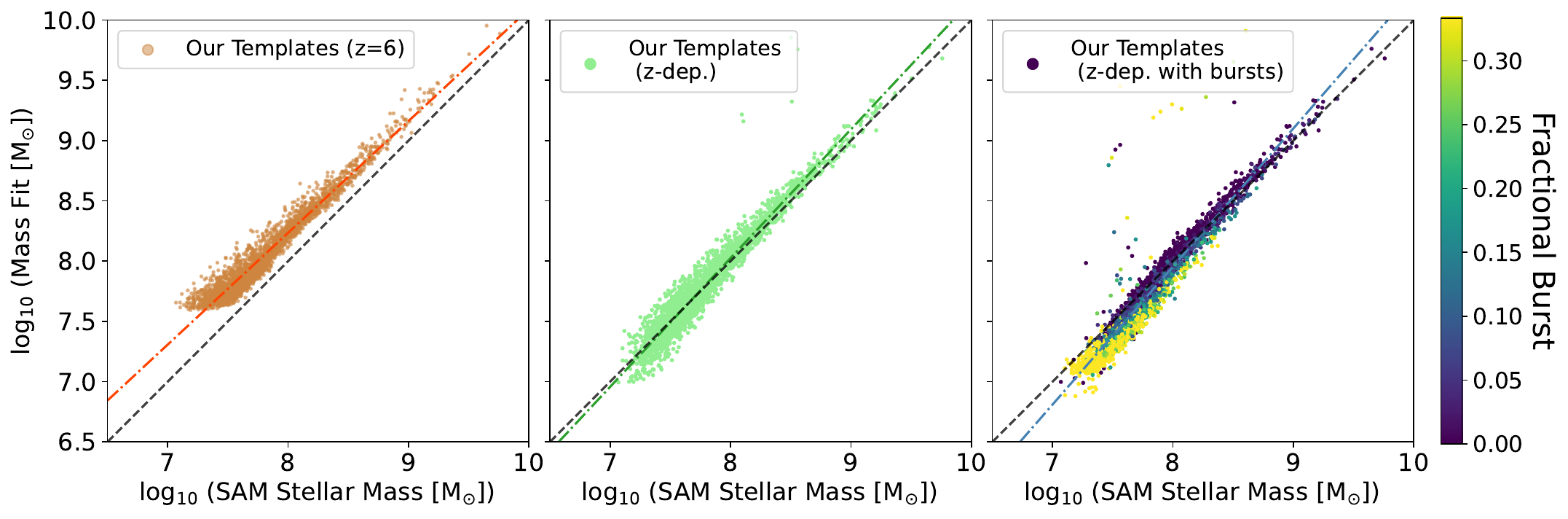}
    \caption{Comparison of recovered stellar masses of Santa Cruz SAM galaxies at $z=8$ across different implementations of our template set. We show the individual model galaxies (points) as well as a least-squares, power-law fit to the data (without errors; dot-dashed lines) to make the trend of each template set clear. \textit{Left:} We first create a template set with \textsc{ares} galaxies but only using spectra from $z = 6$. Because $z=6$ galaxies have longer SFHs and more red stars than higher-$z$ galaxies, these templates systematically overestimate the stellar masses. 
    \textit{Middle:} We then made a template set using \textsc{ares} galaxies with the same smooth star formation histories, but incorporating redshift dependence. Here, the upward bias in the stellar mass findings has disappeared compared to the $z=6$ template set. \textit{Right:} Here we incorporate the burst templates as well. The resulting fits have a small systematic offset from the Santa Cruz SAM. To explore the origin of this offset, we color-code the galaxies by their level of burstiness.}
    \label{fig:ares_template_comparison}
\end{figure}

We now examine the mass fits of simplified template sets to demonstrate the necessity of a more complex and physically-motivated set. We focus on the stellar mass because it is the most difficult ``basic'' parameter to constrain --- the SFR, for example, is almost always well-measured thanks to the rest-UV light. We first set up a basic template set composed only of \textsc{ares} $z = 6$ galaxies without burstiness. We do not expect this to result in accurate parameters, because galaxies evolve rapidly across the Cosmic Dawn (see figure~\ref{fig:template_properties}) and because bursts may be significant. Indeed, in the left panel of figure \ref{fig:ares_template_comparison}, the stellar mass fits are biased high by a factor of $\sim$2. The power-law fit to the data (without errors; dot-dashed line) shows the bias more clearly. The next improvement to the template set would be to include $z$-dependence (see section~\ref{sec:zdep}). Suddenly, in the middle panel, the upward bias has disappeared. The bias occurred in the left panel because the rest-UV is most constrained in the fitting, so the SFR is measured best. But, at a fixed SFR, the stellar mass increases with cosmic time in the model galaxies. By only considering $z=6$, we thus overestimated the true stellar masses

Finally, in the right panel we include burstiness in our templates (see section~\ref{sec:burst}). As expected, the burstiness template set does not massively change the results. Including the bursts narrows the distribution slightly but also introduces a small systematic bias. Here we also color-code the points with the fractional burstiness of the galaxy, which we define to be the ratio of the mass formed in the burst to the entire mass of the galaxy. While the galaxies with only small contributions from the burst templates straddle the line where the fit and SAM stellar masses are equal, galaxies that have experienced large recent bursts deviate from the line. This may be due to the differences between how the Santa Cruz SAM prescribes burstiness within the model and how burstiness is added into our template set. We examine the treatment of bursts more thoroughly in section~\ref{sec:systematics}.

\subsection{Comparing SFHs Between Models}

\begin{figure}
    \centering
    \includegraphics[width=\textwidth]{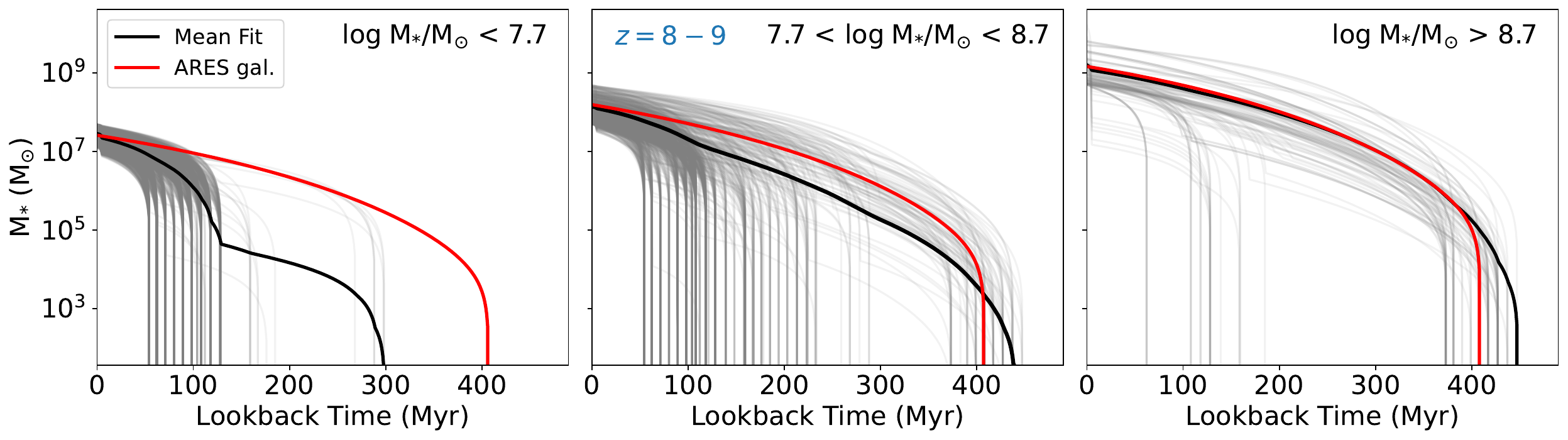}
    \caption{The fitted stellar-mass growth histories for the $z = 8$--$9$ Santa Cruz SAM galaxies from \cite{Yung2022}, organized into three mass bins. The fitted SFHs are in gray. The mean of the fits is in black to showcase the average galaxy at this time. We compare this to the \textsc{ares} galaxy (red) with a stellar mass nearest to the average fitted stellar mass of the JWST galaxies, at the average redshift. Note the general agreement across the two higher mass bins. The lowest mass bin begins forming stars later than the \textsc{ares} galaxy, as expected considering the Santa Cruz SAM has a finite resolution, and cannot resolve stellar mass formed below a certain level (thus the SEDs reflect this lack of SF when the galaxy is below that mass).}
    \label{fig:sfh_fit_sc_sam}
\end{figure}

A particularly useful result from our fitting method is the full, fitted SFH of each galaxy. As explained in section~\ref{sec:full-sfh-model}, because we know the SFH of each template spectrum, the EAZY fits can be turned into an estimated SFH. This is useful to test the agreement between what \textsc{ares} provides as the growth history for a galaxy of a given mass and redshift and the Santa Cruz SAM results. We emphasize that, although \textsc{ares} \emph{expects} all galaxies of a given mass to have the same SFH, the fitting procedure can still test for deviations from that expected history because we allow each source to be fit to galaxies of \emph{any} mass at their redshift. Fits to lower-mass templates would suggest SFHs that begin later and proceed more rapidly, for example. 

To this end, we extract all the Santa Cruz SAM galaxies at $z=8$--$9$ and separate them into three bins based on their stellar masses ($M_{*} < 5 \times 10^{7} \, \mathrm{M_{\odot}}$, $5 \times 10^{7} \, \mathrm{M_{\odot}} < M_{*} < 5 \times 10^{8} \, \mathrm{M_{\odot}}$, and  $5 \times 10^{8} \, \mathrm{M_{\odot}} < M_{*}$). We plot the fitted stellar-mass growth histories for each galaxy in figure \ref{fig:sfh_fit_sc_sam}. Because each of these galaxies are located at a different redshift in the simulation, we plot their SFH as a function of their particular lookback time on the $x$-axis. The fits themselves are in gray, and the mean of these fits is the solid black line. The \textsc{ares} galaxy with a stellar mass nearest to the average stellar mass of the sample, at the average redshift of the sample, is plotted in red. Differences between the thick curves reflect differences in the SFHs of ``average'' galaxies in the different theoretical models.

Overall, we find good agreement in the two larger mass bins of the \textsc{ares} galaxy and the mean growth history line. This means that both the \textsc{ares} and Santa Cruz SAM models are forming stars at a similar rate (on average) over long periods of time. 

However, at the low-mass end, our fits primarily use templates starting star formation \textit{later} than the ``most similar'' \textsc{ares} galaxy. In particular, most of the fits began star formation very recently (less than 100~Myr before the time of observation), suddenly growing to near the current size. This may be a result of the limited resolution of the Santa Cruz SAM merger trees or of their starburst treatment, as opposed to the smooth growth histories in \textsc{ares}. 

In summary, we find that our templates provide very good fits to an independent theoretical model, matching both the galaxy redshifts and stellar masses with only modest biases. We also showed that the templates are flexible enough to capture differences in the growth histories, both on average and for individual systems.

\begin{figure}
    \centering    
    \includegraphics[width=0.6\textwidth]{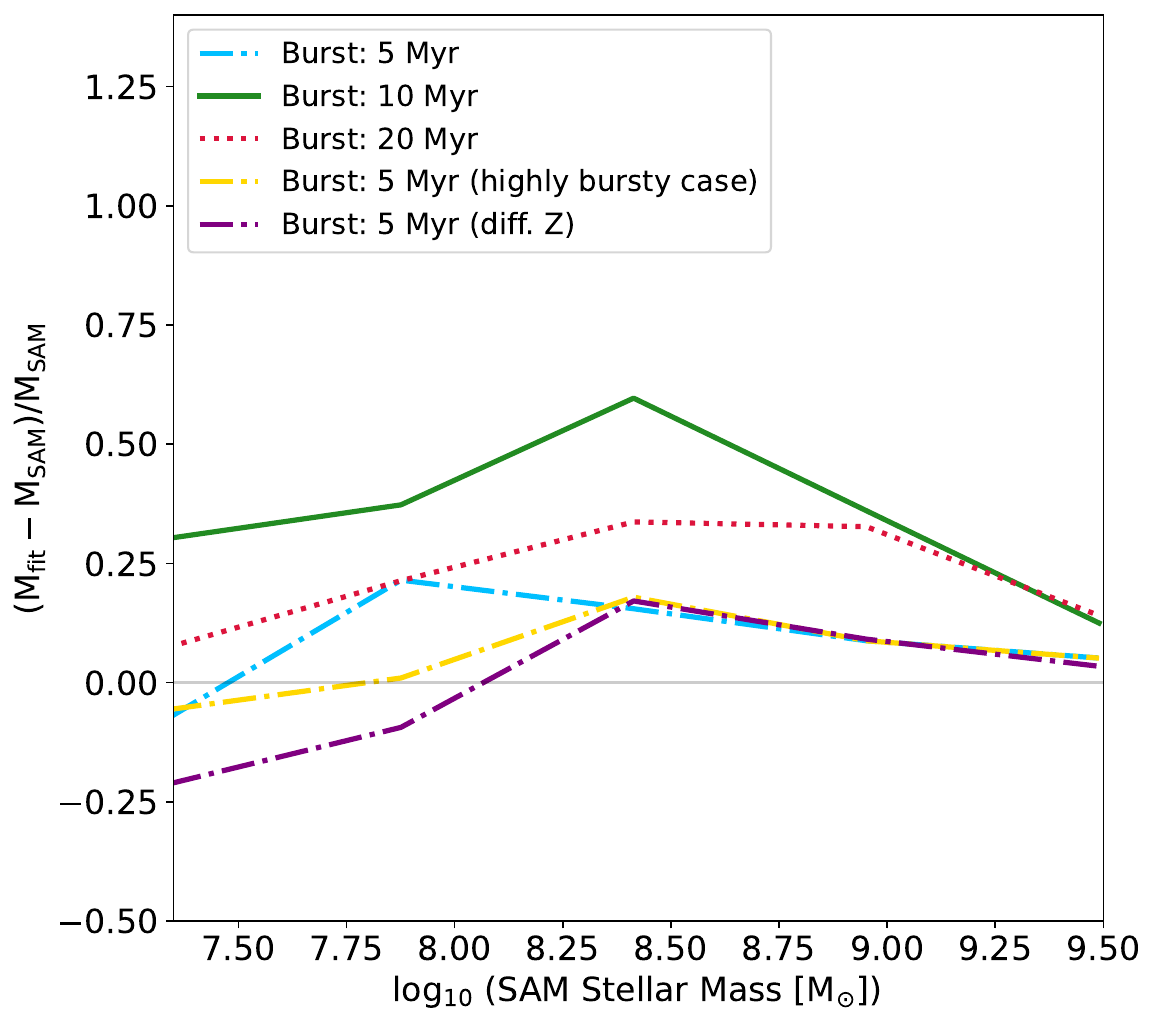}
    \caption{Illustration of some of the systematic effects in our approach. We compare the fitted stellar mass to the true value of the Santa Cruz semi-analytic galaxy sample; for clarity we have plotted the median of each mass bin for our templates. Most of the lines show how the stellar masses differ when we change the assumed age of the burst (our fiducial choice is 5~Myr). In all these cases, we cap the size of bursts to have no more than $\sim$$33 \%$ of the total template mass. The yellow dot-dashed curve increases the maximum allowed burst size. The purple dot-dashed line instead increases the galaxy metallicity.}
    \label{fig:systematics}
\end{figure}

\subsection{Systematic uncertainties in our model framework} \label{sec:systematics}

One of the important issues with SED fitting is the many assumptions that go into the models, including underlying parameters that are rarely modeled explicitly (like the initial mass function of the stellar population, though see \cite{Steinhardt2023}), inputs from stellar atmosphere models (which differ between libraries like, e.g., BPASS and FSPS \cite{Eldridge2009,Conroy2009,Conroy2010}) and the treatment of nebular emission lines \citep{Ferland1980}, and inputs affecting the SED-fitting process itself (like the functional form assumed for the SFH in parametric approaches or the binning scheme for non-parametric models).

While our \textsc{ares}-based templates offer clear advantages in constraining theoretical models, they also make stronger assumptions about many of these issues than other codes. It is therefore important to understand the systematics induced by our choices. While a rigorous estimate is impossible given the many implicit and explicit parameters (and because their effects are also sensitive to the observing strategy!), our comparison to the independent Santa Cruz SAM does help us estimate the importance of these issues.

Figure~\ref{fig:systematics} shows one aspect of these systematics: how some of our assumptions affect the inferred stellar masses. First, we show how the median inferred stellar masses depend on the age of the burst template. Because the UV luminosity is so sensitive to the age of the burst, one can imagine that the stellar mass produced in a burst is very sensitive to this assumption (5~Myr in our fiducial templates). Additionally, although starbursts are present in the semi-analytic model, our single-age model is of course much more restrictive. We see that our fiducial model (blue dot-dashed curve) reproduces the input stellar mass to $\sim 25\%$, though there is a systematic tendency to modestly overestimate the stellar mass. We note that increasing the age of the bursts tends to further increase the masses (dotted and solid curves), although in a non-monotonic way. 

Interestingly, we find that the assumed burst age also has a small but noticeable effect on the inferred redshift: while a 5~Myr burst template set provides redshifts that are nearly unbiased (as in figure~\ref{fig:SC-SAM-z}), longer ages tend to slightly overestimate the redshift (with results closer to the other high-$z$ template sets). 

The yellow dot-dashed curve in figure~\ref{fig:systematics} changes another aspect of burstiness, the maximum allowed burst size. In our fiducial model, we cap the burst stellar mass at 33\% of the total; in this curve, we increase that to 50\%. Fortunately, this has only a small effect on the results, slightly improving the fits in one bin.

Another potential systematic issue is the set of stellar templates used in constructing the model spectra. Our \textsc{ares} galaxies have assumed metallicities and dust relations (with the latter constrained by fits to the colors of HST galaxies at $z>6$ \cite{Mirocha2020}). In the purple dot-dashed curve, we increase the metal retention fraction in the galaxies from $10\%$ to $40\%$. This has no significant effect at large masses but does decrease the estimated mass in small galaxies by $\lesssim 25\%$. We have also tested the dust prescription but found that it has only a small effect on the inferred stellar masses (largely because the dust content is low in almost all systems anyway). 

Finally, we note that \textsc{ares} uses the BPASS v1 stellar libraries from \cite{Eldridge2009}. The Santa Cruz SAM uses the FSPS models of \cite{Conroy2009, Conroy2010}. The differences in the UV and optical spectra between these models are relatively modest, so they should not add significantly to our systematic uncertainties.

In summary, we find that our framework provides stellar mass fits that can be biased by up to a factor of $\sim$50$\%$ relative to the Santa Cruz semi-analytic model, across a fairly broad range of parameters. We could in principle choose \textsc{ares} parameters that provide a better match to the independent model. However, given that JWST has posed challenges to all existing galaxy models at very high redshift, this does not seem especially helpful. Ultimately, we can improve these errors allowing the parameters used to construct the template set to vary and finding their best fit values by comparison to observed systems, which we will do in the future. 

\section{Applying the Templates to a JWST Survey} \label{sec:fitjwstdata}

We now apply our template set to the 717 high redshift galaxies ($z \sim 8$--$18$) identified by \cite{Hainline2023}. They searched a 125 arcmin$^{2}$ region taken with NIRCam in the JWST Advanced Deep Extragalactic Survey (JADES), supplementing with photometry from the Advanced Camera for Surveys (ACS) from HST when available. They selected the galaxies using EAZY. In particular, they used the EAZY ``1.3'' templates, adding nine additional templates: one dusty galaxy, one emission-line galaxy with high equivalent widths, and seven others generated from simulated galaxy surveys. This is the same modified template set compared with our own in figure \ref{fig:SC-SAM-z} (blue points).

Because these galaxies were selected to be at high redshift (although some interlopers are likely), they provide an excellent starting point for our analysis. However, to ensure that these galaxies are high-$z$, we constructed a template set combining our \textsc{ares} template set and the EAZY-default template set to find the redshifts of the galaxies. We modified the EAZY set to be quasi-redshift dependent in that they only fit galaxies $z < 6$, while our template set fit anything $z > 6$. 714 of the initial 717 galaxies fit to a redshift $z > 6$. We cut the other three as interlopers and continued the analysis, as confident as possible that this dataset contains high-$z$ galaxies. We do note that, because none of our templates include dusty, emission-line galaxies, there still may be a low level of contamination from interlopers. We then extracted the physical parameters of this high-$z$ set using our \textsc{ares} templates; the fit of all 714 galaxies from \cite{Hainline2023} took $\sim$3 minutes.

In the remainder of this section, we show some example applications of our template set for physical inference. We note that these are \emph{not} intended to be sophisticated analyses but to demonstrate the capabilities of our \textsc{ares}-based approach for understanding high-$z$ galaxies. In the future, a more complete accounting of survey completeness, errors, and systematic biases are necessary for careful measurements.

\begin{figure}
    \centering
    \includegraphics[width=\textwidth]{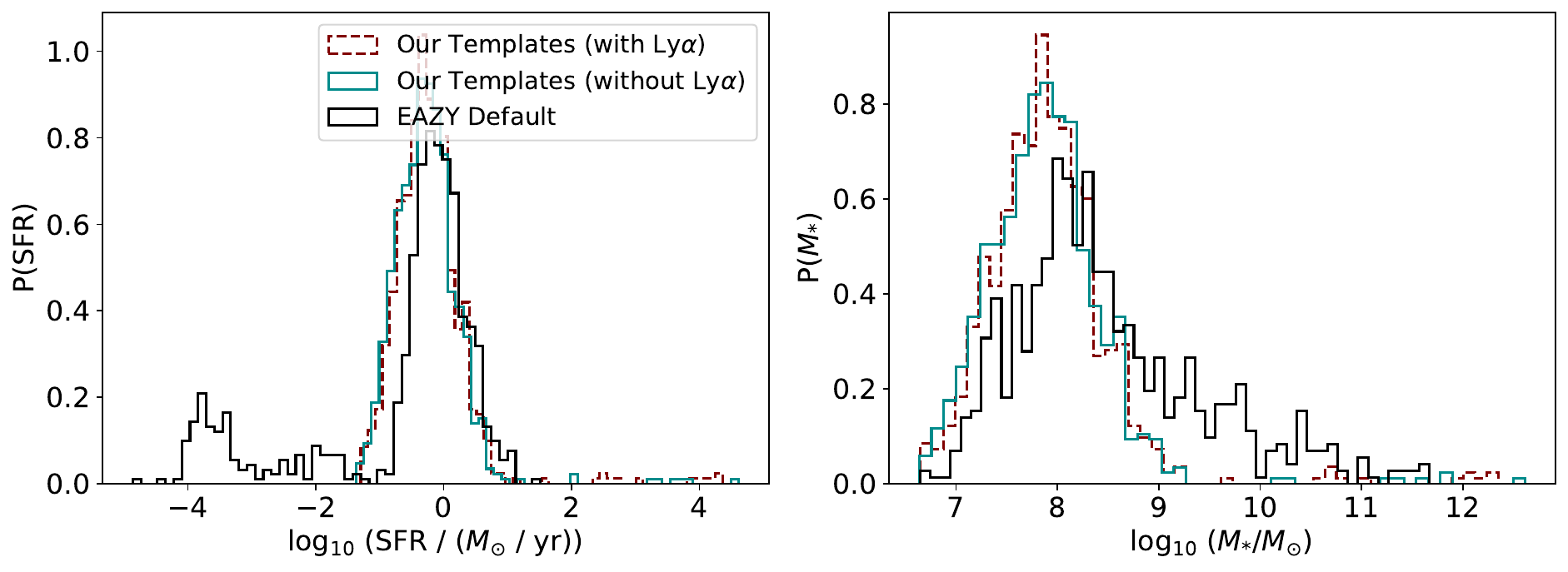}
    \caption{The fitted SFR and stellar mass distributions of the \cite{Hainline2023} high-$z$ galaxies. Each panel includes the results from our template set with and without Ly$\alpha$ (dotted red and solid blue, respectively). Each panel also includes the fit results from EAZY's default template set in black. Note that we have averaged the SFR over the previous 20 Myrs, because some of the galaxies have experienced recent bursts that could obscure the overall trends.}
    \label{fig:hainline_sfr_mass}
\end{figure}

\subsection{Stellar Masses and SFRs}

The most basic results are the distributions of SFRs and stellar masses. We present these in figure \ref{fig:hainline_sfr_mass}. In both panels, our template sets with Ly$\alpha$ are in red and without Ly$\alpha$ are in blue; the EAZY default template set is in black. While we include both choices for Ly$\alpha$ here, we choose the templates without Ly$\alpha$ for the remainder of the analyses. We note, however, that the resulting estimates do not differ significantly.

For the SFR estimates from our template set, we show the time average over the past 20 Myr instead of the instantaneous values, because bursts can skew the distribution and obscure the general trend. We found that a time average of the past 100 Myr did not significantly change the results. Overall, the \textsc{ares} templates return values much closer to the expected range for high-$z$ objects, with larger SFRs but smaller stellar masses. 

Although these distributions are too primitive to be compared directly to theoretical models (as they are affected by survey completeness and selection cuts), they point to some additional advantages of using a theory-based template set. For example, note that the \textsc{ares} results fall off much more steeply with stellar mass than those from the EAZY defaults. This occurs because all of our templates are rapidly star-forming galaxies whose stellar masses increase rapidly with redshift (as expected for average objects at high redshifts). Thus it is very difficult for galaxies to have a large population of old stars dominating their stellar mass (although we will see later that we \emph{can} find evidence for enhanced early star formation). 

This is essentially an implicit prior in our template set against massive, red galaxies at high redshift. While priors can of course be dangerous, if they ``blind'' the analysis to some measurements for instance, they are also advantageous in preventing overfitting. A recent example is highlighted by \cite{Papovich2023} and \cite{Wang2024-massive}. They showed that many galaxies suggested by standard SED-fitting codes to be extremely massive, based on NIR measurements, turned out to be much more reasonable once MIRI measurements were included. Because normal SED-fitting codes do not usually place strong explicit limits on the input parameters, they have a very wide ``prior volume'' that permits a wide range of ``anomalous'' galaxies. Our theory-based templates bake in the assumptions of conventional galaxy formation models, restricting the prior volume in physically-relevant ways that should --- so long as the models are trustworthy --- highlight more conventional interpretations. 

\begin{figure}
    \centering   \includegraphics[width=\textwidth]{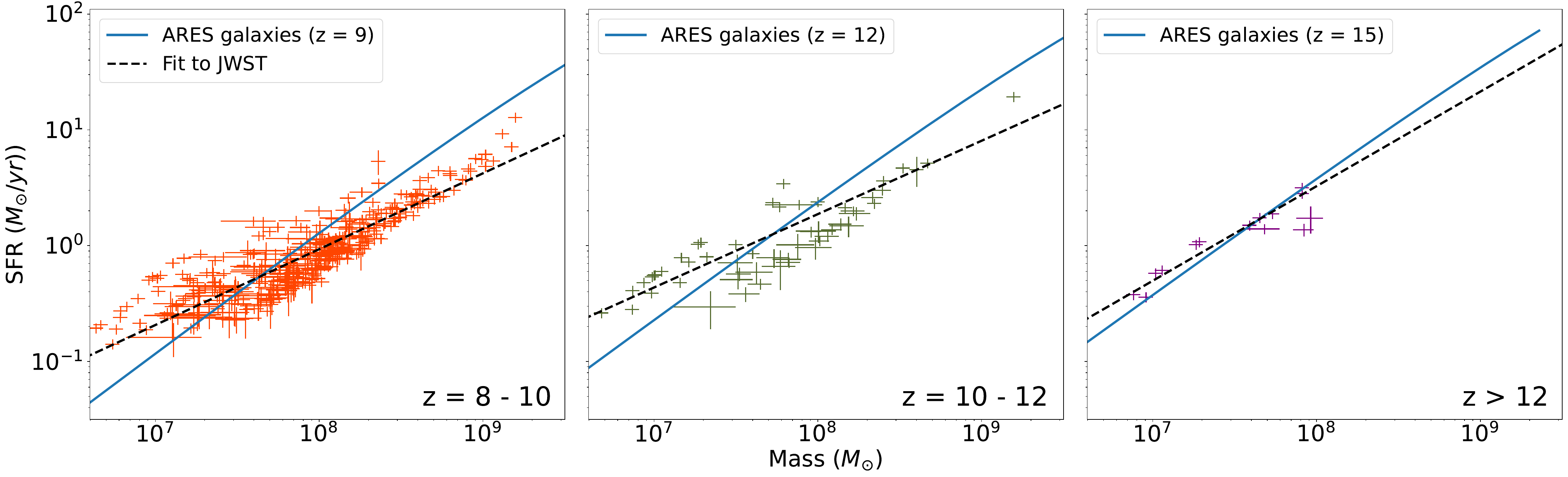}
    \caption{The star-forming "main sequence" inferred from fits to the \cite{Hainline2023} data, split into redshift regimes of $z = 8$--$10$ (left), $z = 10$--$12$ (middle), and $z > 12$ (right). In the $z = 8$--$10$ bin, we have also included the relation found by \cite{Roberts-Borsani2024}, using galaxies from $z = 5$--$7$. By way of comparison, we have plotted the star-forming main sequence curves for the \textsc{ares} galaxy population used to create the template set. There is more agreement as $z$ increases, supporting the need for a deeper analysis in galaxy model parameters.}
    \label{fig:sfms_hainline}
\end{figure}

\subsection{The Star-Forming Main Sequence During the Cosmic Dawn}

We next examine the star-forming ``main sequence'' with this data. We first split the galaxies into three redshift groups: $z = 8$--$10$, $z = 10$--$12$, and $z > 12$. Within each group, we find the stellar mass and SFR from the EAZY fit using the template set without Ly$\alpha$, where in this case the SFR has been averaged over the last 20 Myr. Next, we compute the errors for each data point. EAZY provides the 2.5, 16, 50, 84, and 97.5 percentiles of the probability distribution of each parameter for each galaxy, but these percentiles are not necessarily Gaussian.  In particular, if the $P(z)$ distribution for a galaxy is not Gaussian itself and has a significant lower-$z$ solution as well, the reported uncertainty does not represent the standard deviation and is therefore not meaningful for our desired measurement.

In order to simplify our analysis and ensure our errors are approximately Gaussian, we therefore restrict ourselves to galaxies that we can confidently place at high redshifts. In each redshift group, we cut out galaxies whose redshift 15th percentile is more than $\Delta z = 1$ away from its median. This removes any potential for bimodality in the redshift solution which can cause non-Gaussian errors. To confirm our errors may indeed be approximated as Gaussian, we check the asymmetry between the upper and lower stellar mass and SFR error bars and find them to be small ($< 10 \%$). We also check the asymmetry between the most likely stellar masses and the median stellar masses in EAZY and find them to be also $< 10 \%$. Furthermore, if the errors were less than $10 \%$ of the stellar mass or SFR, we set them to a floor of $10 \%$, motivated by our consideration of systematics in Section~\ref{sec:systematics}.

Knowing that these galaxies' stellar mass and SFR error bars are approximately Gaussian, we plot the star-forming main sequence in figure \ref{fig:sfms_hainline}. The left, middle, and right panels show our redshift bins of $z = 8$--$10$, $z = 10$--$12$, and $z > 12$, respectively. We find a relation between the stellar mass and SFR in log-log space by finding the best fit power law for each group. The fits can be written as 
\begin{equation}
\mathrm{log}_{10} \; (\mathrm{SFR}/1 M_{\odot} \ \mathrm{yr}^{-1}) = a \; \mathrm{log}_{10} \; (M/ 10^{8} M_{\odot}) + b
\end{equation}
where $a = (0.65 \pm 0.02, \, 0.63 \pm 0.04, \, 0.82 \pm 0.07)$ and $b= (-0.03 \pm 0.01, \, 0.27 \pm 0.03, \, 0.51 \pm 0.05)$ for $z=(8$--$10, \, 10$--$12, \, >12$), respectively. 

The increase in the SFR at $10^{8} M_{\odot}$ of the main sequence from $z = 8$ to $z > 12$ follows the physical intuition that, at a given stellar mass, the SFR must increase at higher redshifts in order to build the galaxy in a shorter time frame. It also matches the theoretical expectation that dark matter halos grow more rapidly at early times (e.g., \cite{Dekel2013}). 
In the leftmost panel we also plot the relation found by \cite{Roberts-Borsani2024} for $z = 5$--$7$ galaxies, which is also consistent with this trend. The slope at $z<10$ is consistent with the lower redshift relation as well, although we find it steepens at higher redshifts (albeit with few datapoints and thus large errors).

We also plot the star-forming main sequence curves for the \textsc{ares} galaxy population used to create the template set (solid curves). Interestingly, the \textsc{ares} relation is significantly steeper than the best-fit curve, at least at lower redshifts. The \textsc{ares} galaxies have a very nearly linear relation between the SFR and stellar mass, consistent with measurements of the main sequence at lower redshifts (e.g., \cite{Popesso2023}), and implying a roughly constant specific star formation rate for galaxies at a fixed redshift. The shallower relation implied by our data means that smaller mass galaxies are forming their stars more quickly --- as we will examine below. There is more agreement between the JWST fits and the \textsc{ares} data as redshift increases, although the number of datapoints is quite small in the higher redshift bins. 

This discrepancy between the steeper \textsc{ares} model and the shallower star-forming main sequence fit is both good and bad news. Obviously, it implies that the model used to generate our spectral templates does not accurately describe galaxy formation during this era. But the mismatch is comforting because it demonstrates that the priors of the model do not dominate the results. A wide range of growth histories \textit{may} be fit by superposing the templates corresponding to different galaxy masses at each redshift. The mismatch can then be used to diagnose how the model must be updated. In the present case, burstiness appears to be particularly important --- which is not included as a self-consistent element of our models, so we defer a closer look to future work.

\begin{figure}
    \centering
    \includegraphics[width=0.6\textwidth]{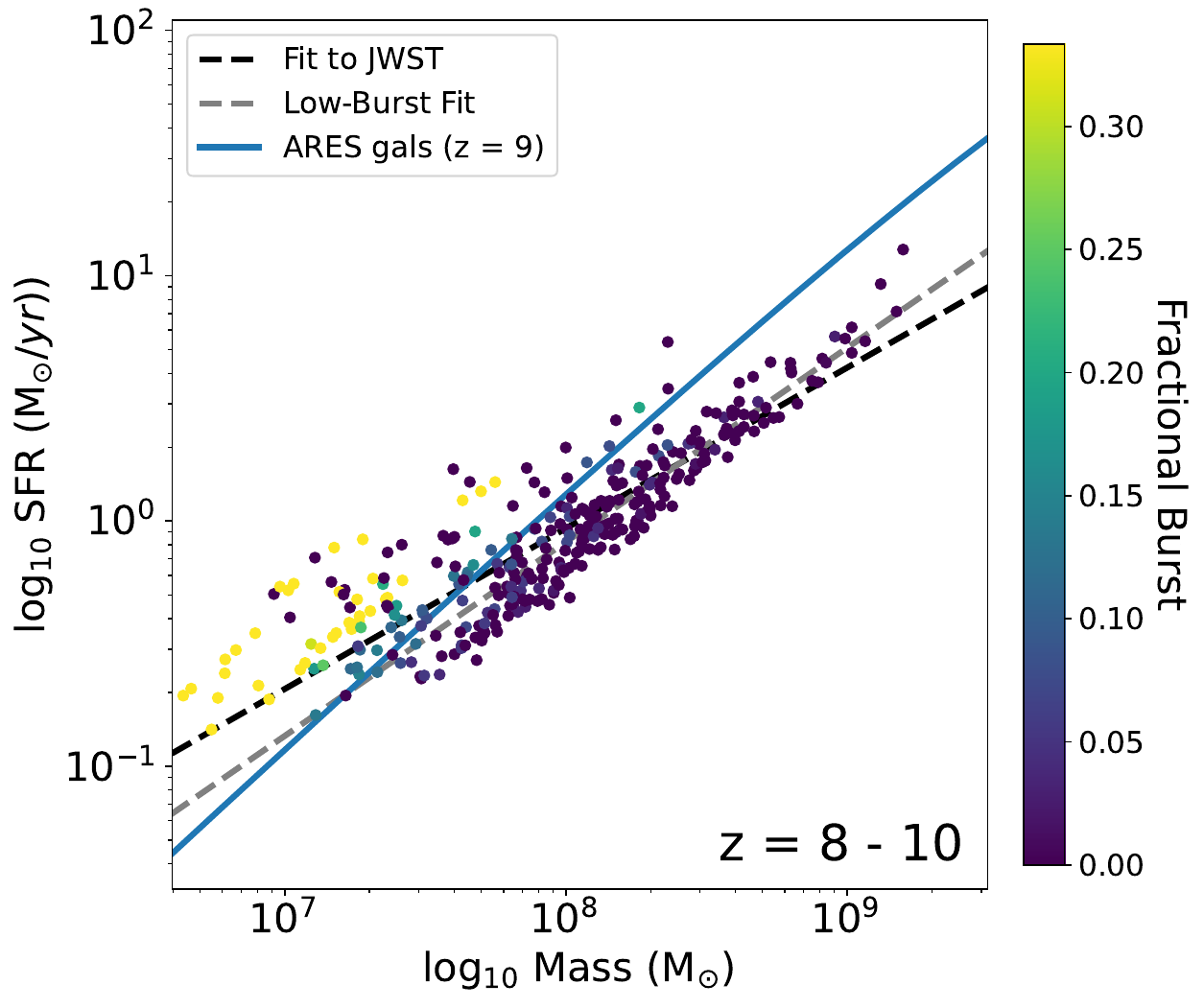}
    \caption{The star-forming main sequence for galaxies at $z=8$--10, color-coded by the amount of burstiness they experience. We see that galaxies with recent bursts scatter above the quiescent galaxies, as expected. We then fit a line (black, dotted) to the galaxies who were inferred to be without bursts and compare it with the total fit (grey, solid) and the \textsc{ares} population (blue, solid).}
    \label{fig:sfms-burst-8-to-10}
\end{figure}

Interestingly, it is obvious by eye in figure~\ref{fig:sfms_hainline} that there is a bimodality to the galaxy distribution amongst the $z=8$--$10$ and $z=10$--$12$ galaxies. In figure \ref{fig:sfms-burst-8-to-10}, we show the main sequence again but color-code the $z = 8$--$9$ sources by their fractional burstiness. The normalization of the main sequence varies with the strength of the recent burst, which causes the bimodality. Motivated by this, we show the fit to the galaxies who were inferred to be without bursts as a black, dotted line.\footnote{We do not fit the high-burst part because it is dominated by the single 5~Myr burst, which has a fixed SFR-stellar mass relation built in.} The low-burst case is steeper and closer to the \textsc{ares} expectations (which, recall, do not include bursts). 

\begin{figure}
    \centering
    \includegraphics[width=\textwidth]{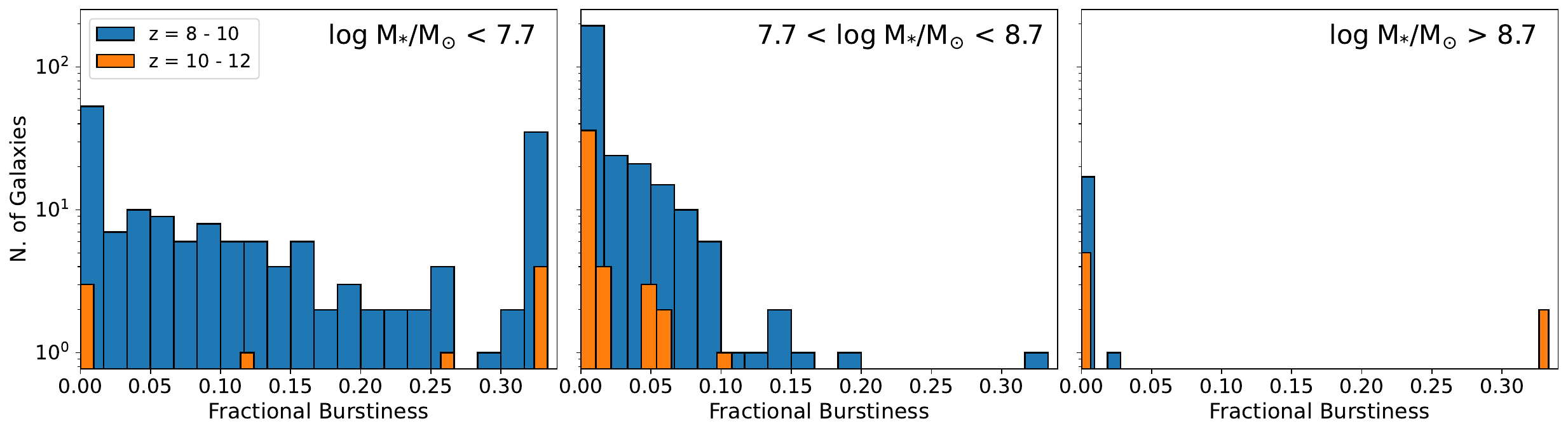}
    \caption{Histograms of the estimated ``burstiness'' of the galaxies from the \cite{Hainline2023} sample split into mass bins of $M_{*} < 5 \times 10^{7} \, \mathrm{M_{\odot}}$, $5 \times 10^{7} \, \mathrm{M_{\odot}} < M_{*} < 5 \times 10^{8} \, \mathrm{M_{\odot}}$, and  $5 \times 10^{8} \, \mathrm{M_{\odot}} < M_{*}$. Within each mass bin, we plot the galaxies between $z = 8$--$10$ in blue and $z = 10$--$12$ in orange. Here burstiness is defined as the ratio of mass formed during the burst to the total stellar mass of the galaxy. We find that burstiness appears to decrease with stellar mass, at least at $z<10$. Trends with redshift are difficult to extract because there are only 9 and 7 galaxies in the lowest and highest mass bins at $z>10$.}
    \label{fig:hainline_perc_burst}
\end{figure}

\subsection{Burstiness of High-$z$ Galaxies}

The burstiness experienced by high-$z$ galaxies is useful for constraining galaxy formation. Motivated by the new measurements of high-$z$ galaxies by JWST, recent efforts have been made to understand the burstiness of SF at high-$z$ \citep{Dekel2023, Pallottini2023, Sun2023a, Asada2024, Clarke2024, Ciesla2024}. Knowing when galaxies transition from bursty star formation to smooth star formation, from one phase to the other, tells us about the growth of galaxies and the processes driving accretion and star formation.  With our template set, we can estimate the burstiness of each galaxy and also the distribution of burstiness across the galaxy population.

Figure~\ref{fig:hainline_perc_burst} shows the resulting distribution, split into three mass bins, with $z = 8$--$10$ galaxies in blue and $z = 10$--$12$ galaxies in orange. We define the burstiness of a galaxy as the fraction of stellar mass formed in a recent burst compared to the total mass (including both the \textsc{ares} galaxy and the burst). This is provided directly by our template fits through the amplitudes of the templates with and without bursts. 

Figure~\ref{fig:hainline_perc_burst} allows us to begin to explore how burstiness may vary with galaxy properties. Theoretical models are unsettled on the expected trends --- \cite{Furlanetto2022}, for example, argues that bursts will be occur in small galaxies, while \cite{Dekel2023} argue that they will appear in larger systems at high redshifts. Our measurements show that, at $z=8$--$10$, burstiness is much more prominent in small galaxies. Unfortunately, the sample at $z>10$ is too small (with only 9 and 7 galaxies in the lowest and highest mass bins) to draw any significant conclusions. We also note that very small galaxies at $z > 10$ have only very recently begun star formation, so the 5 Myr burst may not be as differentiable from the standard galaxy spectrum at this time. Still, this plot shows that our approach can be applied to the archaeology of starbursts to elucidate their origin with larger surveys.

There also can be concern about the edges of the prior on burstiness. There is a pileup at the most-bursty end of the plots, especially in the low-mass, $z = 8$--$10$ distribution. Here, $\sim$20$\%$ of the galaxies have piled up in the bin with maximum allowed burstiness (33\% of the stellar mass in the burst), whereas in the middle and high-mass bins, only one and zero $z=8$--$10$ galaxies are at the maximum limit, respectively. If we increase the maximum burst of the template set from $50 \%$ of the \textsc{ares} galaxy mass to $100 \%$ of the \textsc{ares} galaxy mass, so that it contributes $50 \%$ of the template mass, the percentage of galaxies that pileup at the maximum burstiness in the lowest mass bin at $z = 8$--$10$ is still $\sim$$17 \%$. Evidently, a non-negligible fraction of low-mass galaxies has formed very recently during this time, either through recent galaxy mergers or extreme starburst events.

\begin{figure}
    \centering
    \includegraphics[width=\textwidth]{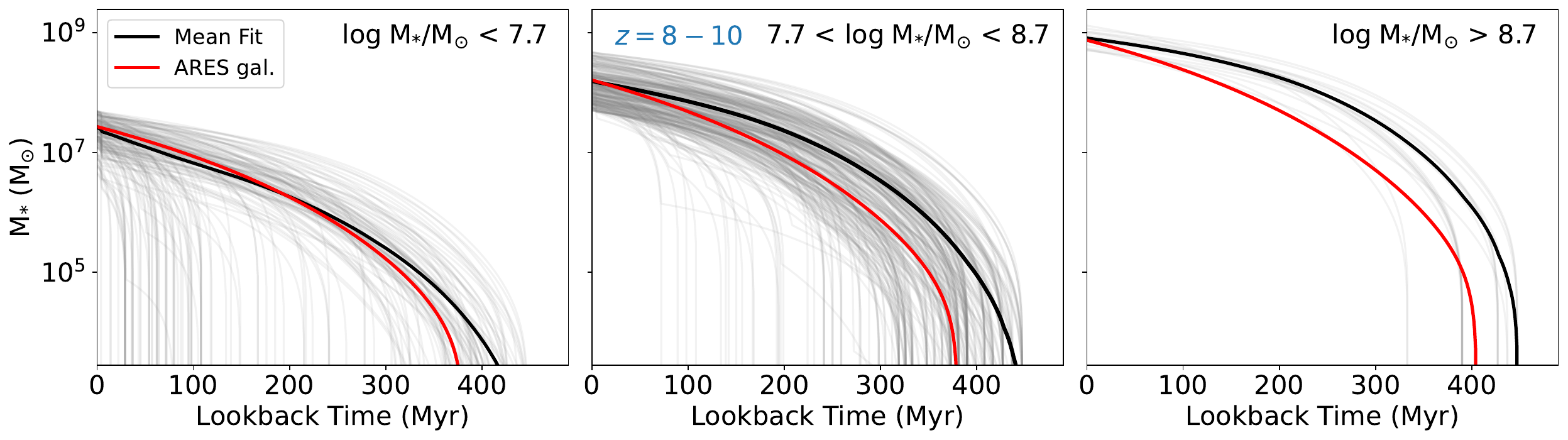}
    \caption{The inferred stellar-mass growth histories of the $z = 8$--$10$ \cite{Hainline2023} JWST galaxies across three bins in stellar mass. The light gray lines show individual galaxy fits, while the thick black line shows the average evolution. To relate the inferred growth histories to our model, we plot the evolution of the \textsc{ares} galaxy that is nearest to the mean stellar mass in each bin at the mean redshift (in red). Note that, on average, the observed galaxies form their stars earlier than the \textsc{ares} population, especially at higher masses. This aligns with the growing consensus that high-$z$ galaxies form stars more quickly and earlier than pre-JWST models predicted.}
    \label{fig:sfh_fit_hainline}
\end{figure}

\subsection{Inferred Stellar Mass Histories of JWST Galaxies} \label{sec:sfh-compare}

The goal of this section is to compare the SFHs of high-$z$ JWST galaxies with model galaxies. We performed a similar exercise with the Santa Cruz SAM galaxies in figure~\ref{fig:sfh_fit_sc_sam}, which helped highlight differences between the theoretical models. That exercise showed that our templates were different enough to capture deviations from the ``expected'' \textsc{ares} SFHs, or in other words that the prior on the preferred shape of the growth history is pretty flexible.

Figure~\ref{fig:sfh_fit_hainline} shows the inferred star formation histories for the JADES galaxies with $z = 8$--$10$ galaxies, split into three mass bins: $M_{*} < 5 \times 10^{7} \, \mathrm{M_{\odot}}$, $5 \times 10^{7} \, \mathrm{M_{\odot}} < M_{*} < 5 \times 10^{8} \, \mathrm{M_{\odot}}$, and  $5 \times 10^{8} \, \mathrm{M_{\odot}} < M_{*}$. The inferred stellar mass histories of the individual galaxies are in gray, and the average of these histories is in black. The \textsc{ares} galaxy at the average redshift of these galaxies, whose mass is closest to the average mass of the observed galaxies, is in red. The differences between the black and red curves hint at population-wide discrepancies between the model and data. We find that in all cases the \textsc{ares} galaxy begins forming stars later than the average observed system (which also means it must grow more rapidly at late times). The difference is most prominent in the highest mass bin (right panel). This suggests that the observed systems require more star formation at very early times than the model predicts. 

This conclusion is consistent with other JWST observations. Detailed SED-fitting of some massive galaxies have shown unexpectedly vigorous star formation in their early history (e.g., \cite{Glazebrook2024, deGraaff2024}). The luminosity function at $z>10$ is also much higher than predicted (e.g., \cite{Harikane2023,PerezGonzalez2023,Donnan2024}). 

\section{Conclusions} \label{sec:concl}

We have created a set of SED-fitting templates tuned to fit high redshift galaxies. In contrast to other custom sets for high-$z$ surveys (e.g., \cite{Larson2023, Hainline2023}), we construct our templates directly from the public galaxy formation code \textsc{ares}. We then incorporate them into the EAZY SED-fitting code \citep{Brammer2008}. By selecting galaxies that span a wide range of masses at each redshift, the templates have a wide range of star formation histories, with the onset of star formation spread across $z \sim 8$ to $z \gtrsim 19$ and diverse SFH shapes and dust content. 
We use an option in EAZY that makes each of these templates ``redshift dependent,'' which results in a correspondence between input galaxies at some redshift $z$ and what \textsc{ares} believes the population of galaxies at redshift $z$ look like. Finally, we have made our templates ``bursty,'' up to a maximum burst size of pre-burst $50 \%$ the stellar mass. The final templates are publicly available at \url{https://github.com/JudahRockLuberto/high-z_template_set}.

Because our templates' star formation histories are determined directly by the model, SED-fitting immediately yields an estimate of not just the redshift but also the SFR, stellar mass, and even past star formation history. However, these conclusions are heavily informed by the model. To validate our approach, we have applied our framework to a completely independent theoretical model, the Santa Cruz semi-analytic model presented in \cite{Yung2022}. We found: \emph{(i)} our template set measures redshifts with a similar precision to other high-$z$-oriented template sets (figures~\ref{fig:SC-SAM-z} and \ref{fig:SC-SAM_chi2}); \emph{(ii)} redshift-dependent templates are essential to accurate fits, because galaxies evolve so rapidly over this epoch (figure~\ref{fig:SC-SAM-EAZY-only}); \emph{(iii)} the properties inferred with our template set  are more accurate than the EAZY default template set (figure~\ref{fig:SC-SAM-EAZY-only}); and \emph{(iv)} although the templates are constructed with a specific theoretical model, they span a diverse enough set of SFHs that the inferences are not dominated by the priors of that model (e.g., figure~\ref{fig:sfh_fit_sc_sam}).

Confident in the robustness of our fit framework, we then applied them to a sample of $\sim$700 galaxies from the JADES survey \citep{Hainline2023}. We found:
\begin{enumerate}
    \item The typical observed galaxy in this sample has an SFR~$\sim 1 \ M_\odot$~yr$^{-1}$ and $M_\star \sim 10^8 \ M_\odot$ (figure~\ref{fig:hainline_sfr_mass}). There are many fewer outliers to small SFR or large stellar mass in our model than with the default EAZY template set. 
    \item The observed sample has a flatter star-forming main sequence than the \textsc{ares} model (figure~\ref{fig:sfms_hainline}). Part of the discrepancy is likely due to burstiness, which pushes many of the smaller galaxies to larger SFRs (figure~\ref{fig:sfms-burst-8-to-10}). 
    \item High-$z$ galaxies observed with JWST begin forming stars earlier (and more rapidly) than our models predict (figure~\ref{fig:sfh_fit_hainline}). 
    \item Burstiness appears prevalent in smaller mass galaxies at $8 < z < 10$, although this trend is uncertain given the small numbers of galaxies at larger stellar masses (figure~\ref{fig:hainline_perc_burst}). However, our template set has the ability to measure trends in burstiness with mass and redshift in a larger sample.
\end{enumerate}

Finally, we highlight three key advantages of our framework. First, in comparison to most SED-fitting codes that construct parameterized or non-parametric star formation histories from simple stellar populations, our SED fitting is orders of magnitude faster (taking only a fraction of a second per source). EAZY has long been used for rapid redshift estimation; our templates allow physical parameters to be measured simultaneously.

Second, in comparison to other phenomenological high-$z$ EAZY templates (e.g., principal component-esque template spectra, templates of the average high-$z$ galaxies, templates which span color space), our framework adds the capability to measure physical parameters via the \textsc{ares} model. 

Third, our framework allows \emph{direct} comparisons between data and models. The phenomenological approach can provide good fits, but unfortunately there is no feedback loop that can point toward necessary improvements in the underlying galaxies. For example, if fits suggest much more massive galaxies than expected, how can we disentangle problems with the fits from problems with galaxy formation models? By coupling the galaxy formation models directly to the observations, our framework points toward the necessary model improvements (including burstiness and earlier formation of massive galaxies, as in figures~\ref{fig:hainline_perc_burst} and \ref{fig:sfh_fit_hainline}).

In this paper, we have only begun to explore the possibilities of the model-data connection. For example, in the future we could vary the \textsc{ares} galaxy parameters to optimize the agreement between the templates and the data, which would help us learn about the underlying physics of galaxy formation. 

\section{Acknowledgments}

We thank Kevin Hainline for supplying JWST photometry and Aaron Yung for providing useful information about the Santa Cruz SAM. This work was supported by by NASA through award 80NSSC22K0818 and by the National Science Foundation through award  AST-2205900. J.M. was supported by an appointment to the NASA Postdoctoral Program at the Jet Propulsion Laboratory/California Institute of Technology, administered by Oak Ridge Associated Universities under contract with NASA. Part of this work was done at Jet Propulsion Laboratory, California Institute of Technology, under a contract with the National Aeronautics and Space Administration (80NM0018D0004). This work has made extensive use of NASA's Astrophysics Data System (\href{http://ui.adsabs.harvard.edu/}{http://ui.adsabs.harvard.edu/}) and the arXiv e-Print service (\href{http://arxiv.org}{http://arxiv.org}), as well as the following software: \textsc{matplotlib} \cite{Matplotlib}, \textsc{numpy} \cite{numpy}, \textsc{astropy} \cite{Astropy}, and \textsc{scipy} \cite{Scipy}.


\bibliographystyle{JHEP.bst}
\bibliography{refs}




\end{document}